\documentclass[preprint,nofootinbib,aps,superscriptaddress,eqsecnum]{revtex4-1} 
\pdfoutput=1
\usepackage{rotating}
 \usepackage{graphicx}
 \usepackage{xcolor}
 \usepackage{amsmath}
\usepackage{amsfonts}
\usepackage{array}
\usepackage{subfigure}
\usepackage{amssymb}
\usepackage{hyperref}
\usepackage{gensymb}
\usepackage{mathrsfs} 
\usepackage{slashed}
\usepackage{caption}
\usepackage{ulem}
\usepackage[mathcal]{eucal}
\def\bea{\begin{eqnarray}}
\def\eea{\end{eqnarray}}
\def\be{\begin{equation}}
\def\ee{\end{equation}}

\newcommand{\rr}{\text{r}}

\begin{document}
%\preprint{LIGO-P1800035}
\title{\Large \bf Shadows of a generic class of spherically symmetric, static spacetimes}
%%%%%%%%%%%%%%%%%%%%%%%%
\author{Md. Golam Mafuz}
\email[Email Address: ]{golammafuzgm@gmail.com}
\affiliation{ Department of Physics, Jadavpur University, Kolkata, 700032, India}
%%%%%%%%%%%%
\author{Rishank Diwan}
\email[Email Address: ]{rishank2610@gmail.com}
\affiliation{ Department of Physics, Indian Institute of Technology Kharagpur, 721 302, India}
\affiliation{ Laboratory for Space Research, The University of Hong Kong, Hong Kong}
%%%%%%%%%%%%%%%
\author{Soumya Jana}
\email[Email Address: ]{soumyajana.physics@gmail.com (Corresponding author)}
\affiliation{ Department of Physics, Sitananda College, Nandigram, 721631, India}
\affiliation{ Department of Physics, Indian Institute of Technology Kharagpur, 721 302, India}
%%%%%%%%%%%%%%%%%
\author{Sayan Kar}
\email[Email Address: ]{sayan@phy.iitkgp.ac.in  (Corresponding author)}
\affiliation{ Department of Physics, Indian Institute of Technology Kharagpur, 721 302, India}
%\affiliation{${}^2$ Centre for Theoretical Studies \\ Indian Institute of Technology Kharagpur, 721 302, India}
%%%%%%%%%%%%%%%%%%%%%%%%%%%%%%%%%%%
\begin{abstract}
\noindent We explore the characteristics of shadows for a general class of spherically symmetric, static spacetimes, which may arise in general relativity or in modified theories of gravity. The chosen line element involves a sum (with constant but different coefficients) of integer powers of $\frac{1}{\text{r}}$ in $\text{g}_\text{tt}$ and $\text{g}_\text{rr}$, in the Schwarzschild gauge. We begin our discussion by motivating the line element through a study of the energy conditions (null and weak) and the extent to which they are satisfied/violated for diverse choices of the parameters appearing in the metric functions. Subsequently, we construct the circular shadows and analyse the dependence of the shadow radius on the metric parameters. We find that with specific choices of the metric parameters (within the ranges allowed by the energy conditions) one can, in principle, obtain values that conform with recent observations on shadows, as available in the literature. We also mention where such metrics may arise (i.e., in which theory of gravity and the physical scenario therein), thereby proposing that the observed shadows may be representative signatures of different theoretical contexts. 
\end{abstract}

\maketitle
\section{Introduction}
%In April $2019$, the Event Horizon Telescope \text{(EHT)} collaboration, a \text{VLBI} radio telescope array operating at $1.3 \, \text{mm}$ wavelengths with Earth-wide baseline, released the image of the shadow of a supermassive compact object $\text{M}\, {87}^{*}$, residing at the center of the nearby supergiant elliptical galaxy \text{Messier} $87$ \cite{event2019first,akiyama2019first2,akiyama2019first3,akiyama2019first5,akiyama2019first6}. This opened the possibility of probing a highly dynamical region of strong gravity in the vicinity of a black hole. More recently, similar observational imaging results on the shadow of the supermassive object SgrA$^*$ located in our own galaxy, have provided further motivation for renewed investigations on the physics around black holes.

 \noindent The recent observations of shadows of M87* \cite{event2019first,akiyama2019first2,akiyama2019first3,akiyama2019first5,akiyama2019first6} and SgrA* \cite{EHT2022_1,EHT_SgrAII,EHT_SgrAIII,EHT_SgrAIV,EHT_SgrAV,EHT2022_6} by the EHT collaboration have opened up the new possibility of probing a highly dynamical region of strong gravity in the vicinity of a supermassive black hole/ compact object .  A black hole's shadow \cite{2000ApJ...528L..13F} (also known as a silhouette \cite{broderick2009imaging,dokuchaev2020silhouettes}) is the apparent (i.e., gravitationally lensed) image of the photon sphere (a region of spacetime where photons move along unstable orbits), appearing as a two-dimensional dark zone in the observer's sky \cite{cunha2018shadows, perlicktsupko, dokuchaev2020silhouettes}. Strong gravitational lensing of light rays near the photon sphere is important for distinguishing between the spacetime structures of black holes and other compact objects \cite{PhysRevD.62.084003, Virbhadra:2002ju,PhysRevD.66.103001,PhysRevD.95.064035,Tsukamoto:2022uoz,PhysRevD.99.104040}. Synge's (1966) seminal study defined the angular radius of the photon capture areas bordering the Schwarzschild black hole  \cite{1966MNRAS.131..463S}. Later, Bardeen explored the Kerr black hole's shadow and observed that the spin would deform the shape of the shadow \cite{1973blho.conf..215B}. Luminet (1979) \cite{1979A&A....75..228L} established through simulations that the photon capture radius of a black hole surrounded by a geometrically thin, optically thick accretion disc would appear to a distant observer as a thin emission ring inside a lensed image. Since then, much effort has gone into expanding our knowledge of how to quantify black hole shadows in known spacetimes employing general relativity
 (GR) \cite{PhysRevD.14.3281, PhysRevD.69.064017,abdujabbarov2013shadow,takahashi2004shapes, PhysRevD.80.024042, shipley2016binary, PhysRevD.86.103001, PhysRevD.92.084005}. Furthermore, the notion of shadows
 has also been analysed for black holes in extended gravity theories \cite{Vagnozzi:2022moj, PhysRevD.81.124045, amarilla2013shadow, Atamurotov:2013dpa, PhysRevD.100.024020, cunha2017shadows, Olmo:2023lil}, for other compact objects \cite{Rosa:2022tfv,PhysRevD.106.044031,PhysRevD.108.044021}, for black hole surrounded by dark matter \cite{Konoplya:2019sns}, etc.   Observational results on black hole shadows are being used to validate theoretical/computational findings on the shadow
 radius, shadow profile as well as astrophysical processes in the strong-field regime (see \cite{johannsen2010testing,psaltis2015general,broderick2014testing,mizuno2018current}).  For reviews one may look at \cite{Zakharov:2014lqa} and
 \cite{perlicktsupko}. As mentioned before, one may
 constrain the parameters in the black hole solutions in  different alternative theories of gravity using concrete facts and numbers emerging from 
 shadow images \cite{2013CQGra..30x4003F,eiroa2014strong}. Apart from  the Schwarzschild solution, the Reissner-Nordstrom (RN) \cite{1916AnP...355..106R, 1918KNAB...20.1238N} metric is the next most basic example within the class of metrics to be discussed in this article. 
Articles in \cite{alexeyev2019phenomenological, alexeyev2020shadows} investigated the effect of including a quadratic in the $\frac{1}{\text{r}}$ term as found in the standard RN solution. Moreover, in many alternative theories of gravity, there exist static, spherically symmetric solutions involving a sum (with constant but distinct coefficients) of integer powers of $ \frac{1}{\text{r}}$ in $\text{g}_{\text{tt}}$ and $\text{g}_\text{rr}$, in the \text{Schwarzschild} gauge. Some examples appear in \cite{1974IJTP...10..363H,babichev2017asymptotically} (see also  \cite{kobayashi2019horndeski} for consequences of extensions of this theory)  \cite{myung2019black,mannheim2012making}, \cite{PhysRevD.9.2273,banerjee2020implications,kar2003static}, \cite{kiselev2003quintessence}(see \cite{visser2020kiselev} for a recent revisit on this metric), and \cite{randall1999large,dadhich2000black}. 
The above-mentioned examples of such line elements motivates us
to investigate, in a broad sense, a general static spherically symmetric metric in the Schwarzschild gauge, involving a sum (with constant but 
different coefficients) of integer powers of $\frac{1}{\text{r}^{\text{n}}}$ upto $\text{n}=4$, in $\text{g}_{\text{tt}} = -{\text{g}_{\text{rr}}^{-1}}$. 
We address (i) the violation/satisfaction of the energy conditions 
for the `matter' (assuming $GR-$like equations) required to support these spacetimes and (ii) the shadow profiles for the above static, spherically symmetric geometries. In both (i) and (ii) our main goal is to find how the
choice of the range of values of the metric parameters influence the consequences. We also use current observational/imaging results to
set bounds on the parameters, which, in turn, can constrain the theory
(when other than GR). In the end, we provide a couple of specific results
for metric functions which may contain a sum of $N$ terms (arbitrary but finite $N$).  While discussing the energy conditions, we have
chosen to use the Weak and Null Energy Conditions, abbreviated as WEC and NEC. 
It is assumed at the outset that our generic line elements are `solutions' in General Relativity or any modified theory in which the Einstein or Einstein-like equations $\text{G}_{\alpha \beta} = \kappa \text{T}_{\alpha \beta}^{\text{eff}}$ hold. As a result, the `required matter' that can be found from  $\text{G}_{\alpha \beta}$ must necessarily be examined for the satisfaction/violation of the energy condition inequalities. In a way, we are actually looking at the
`convergence conditions' (timelike/null). What we broadly call 
`matter' in our analysis, may include `other' quantities eg. an ambient scalar (like in
Brans-Dicke theory).
%The ranges of the various metric parameters for which the energy conditions are obeyed or violated are also obtained.
Our article is divided into six sections. Section \ref{section2} introduces the metrics and comments on the parameters that appear in them. Section \ref{section3} is devoted to the energy conditions (both Weak and Null) for the metrics under consideration. Section \ref{section4} delves into the nature of black hole shadows. Following this, in Section \ref{section5}, we investigate the nature of values for the metric parameters, in the context of current shadow observations. In section \ref{section6} we discuss 
 some results on a more general metric (arbitrary but finite $n$). Finally, in Section \ref{section7}, we summarise our findings and discuss potential consequences. Throughout the paper we assume $\text{G}=\text{c}=1$ and the metric signature convention $(-,+,+,+,)$.
 
 %Unless otherwise stated, we assume the gravitational constant $\text{G}$ and the speed of light $\text{c}$ to be unity, throughout this work. The metric signature convention is $(-,+,+,+,)$.

\section{The class of static, spherically symmetric spacetimes}
\label{section2}

\noindent As mentioned before, we will be concerned with a generic line element  of the following form:
\begin{equation}
    \begin{aligned}\label{metg}
    \text{ds}^{2} = \text{g}_{\alpha \beta} \, \text{dx}^{\alpha} \text{dx}^{\beta}= - \left[ 1+ \frac{\mathcal{P}}{\text{r}} + \frac{\mathcal{Q}}{\text{r}^2} + \frac{\mathcal{S}}{\text{r}^3} + \frac{\mathcal{T}}{\text{r}^4} \right] \text{dt}^{2} +  \left[ 1+ \frac{\mathcal{P}}{\text{r}} + \frac{\mathcal{Q}}{\text{r}^2} + \frac{\mathcal{S}}{\text{r}^3} + \frac{\mathcal{T}}{\text{r}^4} \right]^{-1} \text{d}\text{r}^{2} + \text{r}^{2} \, \text{d} {\Omega_{2}}^{2} 
    \end{aligned}
\end{equation}
where ${\cal P, Q, S, T}$ are constants. We could have kept powers upto say $\frac{1}{\text{r}^\text{N}}$, for a general $\text{N}$. However, it is difficult to say anything concrete for arbitrary $\text{N}$ -- hence our restriction to $\text{N}=4$. Numerous examples of metrics of the above form appear in GR or modified theories of gravity. We mention some of these below in order to motivate our approach and calculations. We will return to some comments on general N in the penultimate section of the article.

\noindent Our first example is the `4-component' generalized Kiselev spacetime \cite{kiselev2003quintessence}  given by the following line element
\begin{equation}\label{metrz}
    \begin{aligned}
    \text{ds}^{2} = - \, \Bigg[ 1- \frac{ \sum_{\text{i}=1}^{4} \, \text{K}_{\text{i}}  \, \text{r}^{-3 \, \text{w}_{\text{i}}}   }{\text{r} }  \Bigg] \,  \text{dt}^{2} +  \Bigg[ 1- \frac{ \sum_{\text{i}=1}^{4} \, \text{K}_{\text{i}}  \, \text{r}^{-3 \, \text{w}_{\text{i}}}   }{\text{r} }  \Bigg]^{-1} \, \text{d}\text{r}^{2} + \text{r}^{2} \, \text{d} {\Omega_{2}}^{2} \,  
    \end{aligned}
\end{equation}
where $\text{d} {\Omega_{2}}^{2}=\text{d}\theta^{2}+\text{sin}^{2}\,\theta\,\text{d}\phi^{2}$ is the line element corresponding to a $2$-sphere. In the above metric, $\text{K}_{\text{i}}$'s are constants containing the parameters appearing in the metrics considered here. We absorb any \text{Schwarzschild} mass term present into one of the $\text{K}_{\text{i}}$ by setting the corresponding $\text{w}_{\text{i}}$ to zero. Thus, essentially, following \cite{jacobson2007gttgrr,visser2020kiselev}, we define a position-dependent mass function m(r) of the form
\begin{equation}\label{mr}
    \begin{aligned}
   2\, \text{m}(\text{r}) =  \sum_{\text{i}=1}^{4} \, \text{K}_{\text{i}}  \, \text{r}^{-3 \, \text{w}_{\text{i}}}
    \end{aligned}
\end{equation}
Comparing the two line elements in equation (\ref{metg}) and (\ref{metrz}), we have $\text{w}_{\text{i}}$'s as follows: $\text{w}_{1} = 0,  \text{w}_{2} = \frac{1}{3}, \text{w}_{3} = \frac{2}{3}, \text{w}_{4} = 1$, and $\text{K}_{\text{i}}$'s as: $\mathcal{P}=-\text{K}_{1}, \mathcal{Q}=-\text{K}_{2}, \mathcal{S}=-\text{K}_{3}, \mathcal{T}=-\text{K}_{4}$. Originally, the Kiselev spacetime was obtained as a solution in GR for quintessence matter surrounding a black hole. However, the Kiselev-type solutions also exist in other contexts such as GR coupled with nonlinear electrodynamics \cite{PhysRevD.106.064017} and black holes surrounded by fluids in $f(R,T)$ gravity \cite{Santos:2023fgd}.  

\noindent The second example stems from early work  in braneworld gravity. An interesting braneworld solution is the tidal Reissner-Nordstr\"om (tidal-RN) braneworld black hole solution \cite{dadhich2000black, whisker2008braneworld}
\begin{equation}\label{Brane}
\begin{aligned}
   \text{ds}^{2} = - \left[ 1 - \frac{2 \text{M}}{\text{r}} \, + \frac{\text{Q}}{\text{r}^2}\right] \, \text{dt}^2 + \left[ 1 - \frac{2 \text{M}}{\text{r}} \, + \frac{\text{Q}}{\text{r}^2}\right]^{-1} \,  \text{dr}^2 + \text r^2 \, \text{d} {\Omega_{2}}^{2} 
\end{aligned}
\end{equation}

\noindent The above solution has the form of the Reissner-Nordstr\"om (\text{RN}) solution of GR with no electric field present on the brane. \text{Q} is a tidal charge parameter arising from the Weyl tensor of the bulk. Unlike in the RN case, \text{Q} can be both positive and negative \cite{dadhich2000black}.

\noindent  The static black hole solutions of the field equations for the quartic Horndeski square root Lagrangian  \cite{babichev2017asymptotically}
constitute our third example. The line element is given as:
\begin{equation}\label{Horndeski}
    \begin{aligned}
   \text{ds}^2 = - \left[1 - \frac{\mu}{\text{r}} - \frac{\beta^2}{2 \zeta \eta \text{r}^2}\right] \, \text{dt}^2 + \left[ 1 - \frac{\mu}{\text{r}} - \frac{\beta^2}{2 \zeta \eta \text{r}^2}\right]^{-1} \, \text{dr}^2 + \text r^2 \, \text{d} {\Omega_{2}}^{2} 
    \end{aligned}
\end{equation}
where $\mu$ is a free integration constant, $\eta$ and $\beta$ are dimensionless parameters, and $\zeta = \frac{ \text{M}_{\textbf{Pl}}}{16 \pi}$. The above solution describes a black hole with mass $\frac{\mu}{2}$. 

\noindent Our fourth example is that of a magnetically charged black hole solution obtained in Extended-Scalar-Tensor-Gauss-Bonnet gravity coupled with nonlinear electrodynamics \cite{PhysRevD.102.104038}. The black hole spacetime is given by
\begin{equation}
    \text{ds}^2= -\left[1-\frac{2M}{r}-\frac{q^3}{r^3}\right]\text{dt}^2+ \left[1-\frac{2M}{r}-\frac{q^3}{r^3}\right]^{-1}\text{dr}^2+ r^2\text{d}\Omega^2_2,
\end{equation}
where $M$ is the mass of the black hole and $q$ is the scalar charge and related to the magnetic charge of the black hole $Q_m=\sqrt{2}q$. The spacetime is completely different from the Riessner-Nordstr\"om- type black holes but falls under the general class that we consider here. The presence of the
$\frac{1}{r^3}$ term in the metric functions is a point to note here. 

\noindent Finally, consider the metric for the black hole surrounded by dust in Rastall theory of gravity given by \cite{HEYDARZADE2017365},
\begin{equation}\label{Rastall}
\begin{aligned}
   \text{ds}^{2} = - \left[ 1 - \frac{2 \text{M}}{\text{r}} \, + \frac{\text{Q}}{\text{r}^2} +    \frac{\text{N}_{\text{d}}}{\text{r}^{\frac{1 - 6 \kappa \lambda}{1 - 3 \kappa \lambda}}} \right] \, \text{dt}^2 + \left[ 1 - \frac{2 \text{M}}{\text{r}} \, + \frac{\text{Q}}{\text{r}^2} +    \frac{\text{N}_{\text{d}}}{\text{r}^{\frac{1 - 6 \kappa \lambda}{1 - 3 \kappa \lambda}}} \right]^{-1} \,  \text{dr}^2 + \text r^2 \, \text{d} {\Omega_{2}}^{2} 
\end{aligned}
\end{equation}
One can realize that in GR, i.e. in the limit of $\lambda \to 0$ and $\kappa = 8 \pi \text{G}$, the black hole in the dust background appears as a charged black hole with an effective mass $\text{M}^{\textbf{eff}} = \text{M} - \text{N}_{\text{d}}/2$. Thus, we see that for $\kappa \lambda \neq 0$, the geometric parameters $\kappa$ and $\lambda$ of the Rastall theory can play an important role leading to distinct solutions relative to GR. One can realize that for  $\kappa \lambda \neq 0$, the Rastall correction term never behaves as the mass or charge terms, and introduces a new character to the black hole.
%, not comparable to the mass and charge terms. 

%It is evident that for $\kappa \lambda \neq 0$, the geometric parameters of the Rastall theory can be crucial in producing different solutions with respect to GR. One can comprehend that the Rastall correction term introduces a new characteristic to the black hole that is clearly different from the mass and charge terms for $\kappa \lambda \neq 0$.

 \noindent We have provided some examples of scenarios in GR or modified gravity where a metric of the form similar to equation (\ref{metg}) arises. There are many more which one can find in the literature. Each of these line elements can be viewed as `solutions' of equations like $\text{G}_{\alpha \beta} = \kappa \, \text{T}^{\textbf{eff}}_{\alpha \beta}$, where the R.H.S. is an effective matter stress energy which contains usual matter plus an `effective' stress energy whose origins could be geometric or anything else. We now move on to discuss the energy conditions of this $\text{T}^{\textbf{eff}}_{\alpha \beta}$ which can be written down from the Einstein tensor for the general line element in equation (\ref{metg}). 

\section{Energy Conditions} \label{section3}

\noindent The \text{Weak Energy Condition (WEC)} \cite{Wald:106274,misner2017gravitation} states that the matter energy density measured by any observer in a spacetime is never negative and we may represent it in the following form
\begin{equation}
\begin{aligned}
\text{T}_{\alpha \beta} \, \text{v}^{\alpha} \, \text{v}^{\beta} \geq 0
\end{aligned}
\end{equation}
where $\text{v}^{\alpha}, \text{v}^{\beta}$ are any future-directed timelike vectors. For a diagonal stress-energy tensor $\text{T}_{\alpha \beta}$, the WEC in a static observer's frame implies
\begin{equation} 
\begin{aligned}
\rho\geq 0 \quad \mbox{and} \quad  \rho + \tau_{\text{i}}\geq 0, \quad \text{i}=1,2,3
\end{aligned}
\label{eq:WEC}
\end{equation}
The \text{Null Energy Condition (NEC)} refers to the following inequality
\begin{equation} 
\begin{aligned}
\text{T}_{\alpha \beta} \, \text{k}^{\alpha} \, \text{k}^{\beta} \geq 0
\end{aligned}
\end{equation}
where $\text{k}^{\alpha}, \text{k}^{\beta}$ are arbitrary, future-directed null vectors. The NEC for a diagonal stress-energy tensor in a static observer's frame corresponds to
\begin{equation}
\quad \rho + \tau_{\text{i}}\geq 0, \quad~ \text{i}=1,2,3
 \label{eq:NEC}
\end{equation}
In Eqs. (\ref{eq:WEC}) and (\ref{eq:NEC}), $\rho$ is the density of mass-energy, $\tau_{1}$ is radial pressure, $\tau_{2}$ and $\tau_{3}$ are tangential pressures. Since the Einstein field equations demand that the stress-energy tensor be proportional to \text{Einstein} tensor, in the static observer's frame the stress-energy tensor $\text{T}_{ \hat{\alpha}  \hat{\beta} }$ should have the same algebraic structure as the Einstein tensor $\text{G}_{ \hat{\alpha}  \hat{\beta} }$. Thus, $\text{T}_{ \hat{\text{t}}  \hat{\text{t}} } = \rho$, $\text{T}_{ \hat{\text{r}}  \hat{\text{r}} } = \tau_{1}$ and $ \text{T}_{ \hat{\theta}  \hat{\theta} } = \text{T}_{ \hat{\phi}  \hat{\phi} } = \tau_{2} = \tau_{3}$, are the only non zero components of the stress-energy tensor \cite{Wald:106274,misner2017gravitation,1988AmJPh..56..395M} into a perfect fluid \cite{boonserm2016mimicking}. The energy conditions (assuming $\text{T}_{\alpha \beta} = \frac{1}{\kappa} \text{G}_{\alpha \beta}$) in the proper reference frame for this metric are the following in-equalities (ignoring the overall positive factor $\frac{1}{\kappa}$):
\begin{equation} 
 \begin{aligned}
\rho = \frac{\text{r}^2 \, \mathcal{Q} + 2\, \text{r} \, \mathcal{S} + 3 \, \mathcal{T} }{\text{r}^6}  \geq 0
 \end{aligned}
 \label{eq:wec}
\end{equation}
\begin{equation} 
 \begin{aligned}
\rho + \tau_{2} = \frac{ 2 \, \text{r}^2 \, \mathcal{Q} + 5 \, \text{r} \, \mathcal{S} + 9 \, \mathcal{T} }{\text{r}^6}  \geq 0
 \end{aligned}
 \label{eq:nec}
\end{equation}

\noindent A crucial point to note here is that $\rho+\tau_{1}= 0$ for all $r$ and all values of the parameters. We also infer the following points from the inequalities:\\
\noindent (a) The inequalities are independent of the parameter $\cal P$.\\
\noindent (b) For ${\cal Q}>0 $, ${\cal S}>0$, ${\cal T}>0$ both inequalities hold for all $\text{r}$. Thus WEC and NEC hold for all $\text{r}$.\\
\noindent (c) For ${\cal Q}>0 $, ${\cal S} <0$, ${\cal T}>0$ both inequalities hold, for all $\text{r}$, if ${\cal{S}}^2 < \frac{72}{25}  \cal{Q} \cal{T}$. This requirement (i.e. $ {\cal{S}}^{2} < \frac{72}{25} \cal{Q} \cal{T}$) follows from the condition for which the roots of the equalities are complex conjugate pairs. One can check that the minimum value of the numerator factor in the R. H. S. of both the inequalities (subject to appropriate choices of ${\cal Q}$, $\cal R$ and $\cal T$) is positive definite. \\
\noindent (d) Other combinations of ${\cal Q}$, ${\cal S}$ and ${\cal T}$ may lead to the WEC, NEC being satisfied over restricted ranges. For example one can write both the numerators as products of the form $(\text{r}-\text{r}_+)(\text{r}-\text{r}_-)$. Among the four values of the roots, if one chooses the largest, then if $\text{r}$ is greater than this largest root, both \text{WEC}, \text{NEC} will hold in that domain of $\text{r}$.

\noindent (e) For ${\cal Q}<0$ the energy conditions are violated for large $r$ irrespective of any values of ${\cal S}$, ${\cal T}$. For ${\cal Q}<0$, ${\cal S}>0$, ${\cal T}<0$ violation of the energy conditions occur for all $r$ when $ {\cal{S}}^{2} < \frac{72}{25} \cal{Q} \cal{T}$, whereas if $ {\cal{S}}^{2} > \frac{72}{25} \cal{Q} \cal{T}$ there exists only the intermediate region of space $r_{-}<r < r_{+}$ where the energy conditions are satisfied. Also, for ${\cal Q}<0$, ${\cal S}<0$, ${\cal T}<0$ the energy conditions are violated at any $r$. 

\begin{figure}[!htbp]
\centering
\subfigure[$Q>0$, $q=0.2$ ]{\includegraphics[width=3in,angle=360]{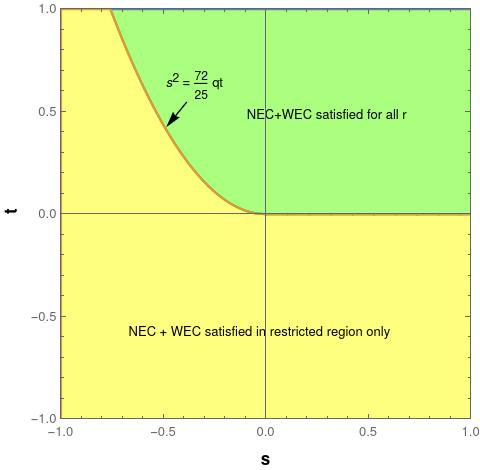}\label{subfig:fAa0}}
\subfigure[$Q<0$, $q=-0.2$]{\includegraphics[width=3in,angle=360]{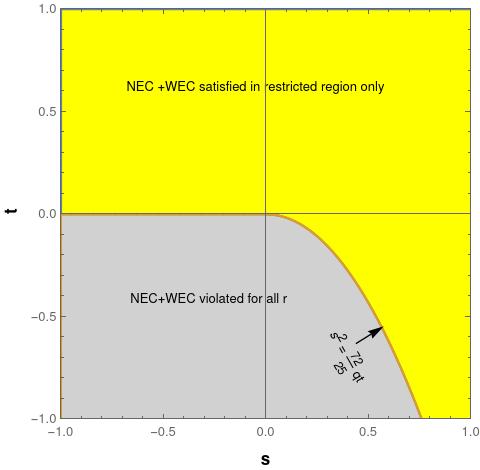}\label{subfig:fAb0}}
\caption{\raggedright $t-s$ region plot (for a fixed $q$) showing the status  of NEC and WEC. In the left panel $(a)$ $q={\cal Q}/M^2>0$. We choose a particular value $q=0.2$ to specify the boundary $s^2=\frac{72}{25}qt$. However, the nature of the plot does not change for any other positive value of $q$. In the right panel $(b)$ $q={\cal Q}/M^2<0$. We choose a particular value $q=-0.2$ }
\label{fig:A0}
\end{figure}

\noindent In Table~\ref{table1}, we summarise the status of NEC and WEC for all possible positive or negative combinations of  $ {\cal Q} $, $ {\cal S} $,  $ {\cal T} $. Fig.~\ref{fig:A0} demonstrates the status of the energy conditions with reference to the available  parameter space.  This status is shown via 
coloured regions
in the plots for $s$ vs. $t$ (where $s={\cal S}/M^3$ and $t={\cal T}/M^4$) for ${\cal Q}>0$ and ${\cal Q}<0$. In the green region, NEC and WEC are satisfied for all $r$. In the yellow region the energy conditions are partially satisfied for restricted $r$, 
whereas the grey region indicates the violation of the energy conditions at all $r$. Thus, the figure provides an overall summary of the details shown in Table~ \ref{table1}.

\begin{table}
\begin{center}
\begin{tabular}{|m{0.5cm}|m{0.5cm}|m{0.5cm}|l|}
\hline
$ {\cal Q} $& $ {\cal S} $ & $ {\cal T} $ & NEC and WEC status \\
\hline
 $+$ & $+$ & $+$ & satisfied for all $r$\\
 $+$ & $+$ & $-$ & satisfied for $r>r_{+,max}$\\
 $+$ & $-$ & $+$ & satisfied for all $r$ when ${\cal S}^2\leq \frac{72}{25}{{\cal  Q}{\cal T}}$; violation occur 
 at $r_{-,min}<r<r_{+,max}$ when ${\cal S}^2 > \frac{72}{25}{{\cal  Q}{\cal T}}$  \\
 $+$ & $-$ & $-$ & violation in between $r_{-,min}<r<r_{+,max}$, satisfied elsewhere \\
 $-$ & $+$ & $+$ & satisfied only for  $r_{-,2nd}<r<r_{+,2nd}$\\
 $-$ & $+$ & $-$ & violation for all $r$ when ${\cal S}^2\leq \frac{72}{25}{{\cal  Q}{\cal T}}$; satisfied only 
 at  $r_{-,2nd}<r<r_{+,2nd}$ when ${\cal S}^2 > \frac{72}{25}{{\cal  Q}{\cal T}}$ \\
 $-$ & $-$ & $+$ & violation occur at $r \geq r_{+,2nd} $  \\
 $-$ & $-$ & $-$ & violation for all $r$ \\
\hline
\end{tabular}
\end{center}
\caption{ \label{table1} \raggedright NEC and WEC validity/violation are tabulated for different signatures of $ {\cal Q} $, $ {\cal S} $,  $ {\cal T} $. Here $r_{+,max}$ and $r_{-,min}$ denote the largest and the smallest among all roots of the numerators of the L.H.S. of Eqs.~(\ref{eq:wec}) and (\ref{eq:nec}) and $r_{+,2nd}$ and $r_{-,2nd}$ are the second largest and second smallest among all the roots. }
\end{table}

\noindent Let us now move on to illustrate the L. H. S. of the energy condition inequalities for different sets of values for the parameters \{$\mathcal{P},\mathcal{Q},\mathcal{S},\mathcal{T}$\}. We need not fix ${\cal P}$ since it does not appear in the inequalities. We will rename ${\cal P}= - 2 \text{M}$. The remaining ${\cal Q}$, ${\cal S}$, ${\cal T}$ are chosen to satisfy the constraints mentioned above. In order to deal with dimensionless quantities we write
\begin{equation}
{\bar r} = \frac{r}{M} \quad~;\quad~ {\mathcal {Q}} = q M^2 \quad~; \quad~ {\mathcal {S}} = s M^3 \quad~ ; \quad~ 
{\mathcal {T}} = t M^4
\end{equation}
where $q,s,t$ are numbers. The two inequalities rewritten in terms of the new variables become
\begin{equation}
\rho = \frac{ q {\bar r}^2 + 2 s {\bar r} + 3 t}{M^2{\bar r}^6} \geq 0
\quad~ ; \quad~
\rho +\tau_2 = \frac{2 q {\bar r}^2 + 5 s {\bar r} + 9 t}{M^2{\bar r}^6} \geq 0
\end{equation}

\begin{figure}[!htbp]
\centering
\subfigure[$q=2$, $s=-1$, $t=3$]{\includegraphics[width=3.2in,angle=360]{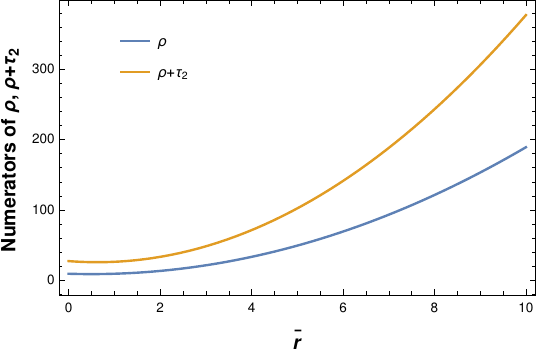}\label{subfig:fAa}}
\subfigure[$q=1$, $s=-2$, $t=1$]{\includegraphics[width=3.2in,angle=360]{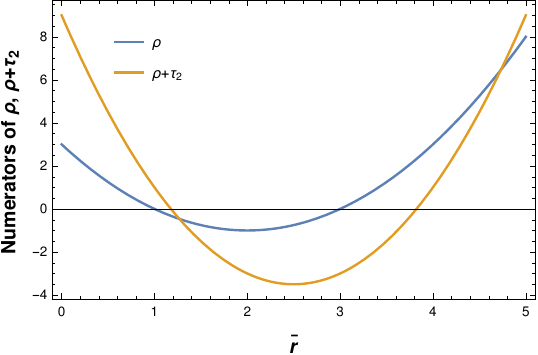}\label{subfig:fAb}}
\caption{\raggedright Numerators of energy condition inequalities: $\rho\geq 0$ inequality
(blue), $\rho+\tau_2\geq 0$ inequality (yellow) are plotted as the function of $\bar{r}$. Note the violation for values not satisfying $s^2< \frac{72}{25} q t$ (figure on the right). }
\label{fig:A}
\end{figure}

\noindent We plot (Figure~\ref{fig:A}) only the
numerators of the above inequalities in order to verify their positivity for all $\text{r}$. The triplet $(q,s,t)$ are chosen as  $(2,-1,3)$ (energy condition satisfying)  and $(1,-2,1)$ (energy condition violating) -- these values are considered with the sole aim of illustration. In general, as mentioned earlier, if required conditions are met, one is sure of satisfying the energy conditions. Later we will try to correlate the shadow radius with scenarios of
metrics for which the required matter satisfies WEC, NEC.

%\begin{figure}[!htbp]
%\centering
%\subfigure[$Q>0$, $q=0.2$ ]{\includegraphics[width=3in,angle=360]{ECregplot1.jpg}\label{subfig:fBa0}}
%\subfigure[$Q<0$, $q=-0.2$]{\includegraphics[width=3in,angle=360]{ECregplot2.jpg}\label{subfig:fBb0}}
%\caption{\raggedright $t-s$ region plot (for a fixed $q$) showing the status  of NEC and WEC. In the left panel $(a)$ $q={\cal Q}/M>0$. We choose a particular value $q=0.2$ to specify the boundary $s^2=\frac{72}{25}qt$. However, the nature of the plot does not change for any other positive value of $q$. In the right panel $(b)$ $q={\cal Q}/M<0$. We choose a particular value $q=-0.2$ }
%\label{fig:B0}
%\end{figure}

%\begin{figure}[H]
%\begin{subfigure}{.5\textwidth}
%  \centering
%  \includegraphics[width=\linewidth]{ec1.eps}  
%  \captionsetup{labelfont={bf,scriptsize},textfont={it,tiny}}
%  \caption{$q=2$, $s=-1$, $t=3$.}
%  \label{ec1}
%\end{subfigure}
%\begin{subfigure}{.5\textwidth}
%  \centering
%  \includegraphics[width=\linewidth]{ec2.eps}  
%  \captionsetup{labelfont={bf,scriptsize},textfont={it,tiny}}
%  \caption{$q=1$, $s=-2$, $t=1$}
%  \label{ec2}
%\end{subfigure}
%\caption{Numerators of energy condition inequalities: $\rho\geq 0$ inequality (blue), $\rho+\tau_2\geq 0$ inequality (yellow). Note the violation for values not satisfying $s^2< \frac{72}{25} q t$ %(figure on the right).}
%\label{fig:1}
%\end{figure}

\section{Analysis of critical null geodesics and the radius of the shadow} \label{section4}

\noindent When light from a distant astrophysical object or the accretion disk surrounding the black hole arrives in the vicinity of the event horizon, a part of it gets trapped inside the horizon while another part escapes to infinity \cite{1966MNRAS.131..463S,1983mtbh.book.....C,PhysRevD.14.3281}. This results in the black hole shadow \cite{2000ApJ...528L..13F} -- the two dimensional closed curve in the observer’s sky separating capture and scattering orbits \cite{1973blho.conf..215B,1979A&A....75..228L, gralla2019black}. However, the shadow may not be a direct consequence of the existence of an event horizon \cite{cunha2018does,abdikamalov2019black}. Rather, it is determined by a set of bound null orbits, exterior to the horizon, dubbed as the `photon sphere' in \cite{perlicktsupko} (light rings in \cite{cunha2018shadows}, photon ring in \cite{2010ApJ...718..446J,johnson2020universal} and critical curve in \cite{gralla2019black}). 

\noindent In this section, we study the null geodesics of photons in a general  static,  spherically symmetric and asymptotically flat line element 
and first work out the equation of the photon sphere. Subsequently, we move on to calculate the radius of the black hole shadow (following Bardeen's formalism)
in terms of celestial coordinates \cite{1973blho.conf..215B, vazquez2003strong, 
2000CQGra..17..123D, cunha2018shadows}. 

\noindent Let us begin with the line element
\begin{equation}
    \begin{aligned}
    \text{ds}^{2} = \text{g}_{\alpha \beta}\, \text{dx}^{\alpha} \,\text{dx}^{\beta} = \, - \mathscr{A}(\text{r}) \, \text{dt}^{2} + \, \mathscr{B}(\text{r}) \, \text{dr}^{2} + \,\text{r}^{2} \, d\theta^{2} + \,\text{r}^{2} \, \text{sin}^{2}\,\theta    \, d\phi^{2}
    \end{aligned}
\end{equation}
 such that, $\underset{\text{r} \to \infty}{\lim} \, \mathscr{A}(\text{r}) = \underset{\text{r} \to \infty}{\lim} \, \mathscr{B}(\text{r})=1$. The Lagrangian describing the motion of photons in this metric is
\begin{equation}
\begin{aligned}
    2 \, \mathscr{L} (\text{x}^{\alpha},{\Dot{\text{x}}}^{\alpha}) =  \text{g}_{\alpha\beta} \, {\Dot{\text{x}}}^{\alpha} \, {\Dot{\text{x}}}^{\beta}
\end{aligned}
\end{equation}
where $\Dot{\text{x}}^{\alpha}$ is the tangent vector along a curve $\text{x}^{\alpha}=\text{x}^{\alpha}(\lambda)$, where  $\lambda$ is the affine parameter. The conjugate momentum corresponding to the coordinate $\text{x}^{\alpha}$ is $\text{p}_{\alpha}$. Since the metric is independent of $\text{t}$ and $\phi$, the corresponding constants of motion are the energy $\text{E}$ and the angular momentum $\text{L}$. By solving these two expressions we can get the geodesic equations corresponding to $\text{t}$ and $\phi$ coordinates.
The Hamiltonian for the photons is
\begin{equation}
\begin{aligned}
   \mathscr{H}= \frac{1}{2} \, \text{g}^{\alpha \beta} \, \text{p}_{\alpha} \, \text{p}_{\beta}=0 
\end{aligned}
\end{equation}
The geodesic equations corresponding to $\text{r}$ and $\theta$ coordinates can be obtained from the Hamilton-Jacobi equation
\begin{equation}
    \begin{aligned}
    \mathscr{H}(\text{x}^{\alpha},\text{p}_{\alpha}) + \frac{\partial \mathscr{S}}{\partial \lambda} = 0
    \end{aligned}
\end{equation}
The Jacobi action $\mathscr{S}$ is integrated with the help of the two known constants of motion and is of the following form
\begin{equation}
    \begin{aligned}
    \mathscr{S} = - \, \text{E} \, \text{t} + \, \text{L} \, \phi + \, \bar{\mathscr{S}}(\text{r},\theta)
    \end{aligned}
\end{equation}

\noindent For the given static, spherically symmetric metric, $\bar{\mathscr{S}}(\text{r},\theta)$ happens to be separable in $\text{r}$ and $\theta$, i.e., $\bar{\mathscr{S}}(\text{r},\theta) = \mathscr{S}^{\text{r}}(\text{r}) + \mathscr{S}^{\theta}(\theta)$. The `$\text{r}$' and `$\theta$' components of the momentum are, respectively,
$\text{p}_{\text{r}}$ and $\text{p}_{\theta}$. Inserting $\text{p}_{\text{r}}$ and $\text{p}_{\theta}$ in the Hamilton-Jacobi equation, and separating out the parts corresponding to $r$ and $\theta$ coordinates, we get
\begin{equation}
    \left[\frac{1}{\mathscr{B}}\left(\frac{\text{d}\mathscr{S}^{\text{r}}}{\text{d}r}\right)^2-\frac{\text{E}^2}{\mathscr{A}}\right]r^2 + \text{L}^2= -\left(\frac{\text{d}\mathscr{S}^{\theta}}{\text{d}\theta}\right)^2- \text{L}^2\cot^2\theta = - \text{K},
\end{equation}
where $\text{K}$ is a separation constant, also known as the Carter constant \cite{PhysRev.174.1559}, representing the third constant of motion. Thus, geodesic equations corresponding to  $\text{r}$ and $\theta$ coordinates are
\begin{eqnarray}
    \mathscr{B}(r) \dot{r}&=&\text{E}\sqrt{\mathscr{B}\left[\frac{1}{\mathscr{A}}-\frac{\ell^2+\chi}{r^2}\right]}=E\sqrt{-V(r)},\\
    \dot{\theta}&=&\frac{\text{E}}{r^2}\sqrt{\Theta(\theta)},
\end{eqnarray}
where $\text{$\ell$}=\frac{\text{L}}{\text{E}}$, $\chi=\frac{\text{K}}{\text{E}^{2}}$ and $\Theta (\theta) = \chi - \text{$\ell$}^2 \, \text{cot}^2 \, \theta$, while the effective potential in which the photons move is given as,
\begin{equation}
    V(r)=\mathscr{B}\left[\frac{\ell^2 +\chi}{r^2}-\frac{1}{\mathscr{A}}\right].
\end{equation}

\noindent The radii of the photon sphere $\text{r}_{\textbf{ph}}$ correspond to the highest maximum of the effective potential $\text{V}(\text{r})$ subject to the following conditions
\begin{equation}\label{cnstreq}
    \begin{aligned}
    \text{V}(\text{r}_{\textbf{ph}}) = 0 , \quad \text{V}^{'}(\text{r}_{\textbf{ph}}) = 0, \quad \text{V}^{''}(\text{r}_{\textbf{ph}}) < 0
    \end{aligned}
\end{equation}
%\sout{The first two of the above conditions give respectively} 
From Eq.(\ref{cnstreq}) we get the following conditions
\begin{equation}
    \begin{aligned}
    \chi + \text{$\ell$}^2 = \frac{ \text{r}^{2}_{{\textbf{ph}}} }{\mathscr{A}( \text{r}_{\textbf{ph}} ) }
    \quad \text{and} \quad 
    \chi + \text{$\ell$}^2 =  \frac {1}{2}  \, \frac{ \mathscr{A}^{'}( \text{r}_{\textbf{ph}}) }{ \mathscr{A}^2( \text{r}_{\textbf{ph}} ) } \, \text{r}^{3}_{\textbf{ph}}
    \end{aligned}
\end{equation}
where a prime denotes derivative w.r.t. the radial coordinate $\text{r}$. We determine the photon sphere radius taking the maximum positive root of the following photon sphere equation.
\begin{equation}
    \begin{aligned}
    \frac{ \mathscr{A}^{'}(\text{r}_{\textbf{ph}}) }{ \mathscr{A}( \text{r}_{\textbf{ph}} ) } \, \text{r}_{\textbf{ph}} = 2
    \end{aligned}
\end{equation}
In order to derive the contour of the black hole shadow in the observer’s sky, one considers the projection of the photon sphere in the image plane. The locus of the shadow boundary is expressed in terms of two celestial coordinates. In terms of the tangent to a photon geodesic at the observer’s position located at a distance $\text{r}_{0}$ from the center of the black hole, the celestial coordinates are \cite{1973blho.conf..215B, vazquez2003strong, 2000CQGra..17..123D, cunha2018shadows}
\begin{equation}
    \alpha= -\frac{\ell}{\sin \theta_0}, \quad~ \beta= \pm \sqrt{\Theta (\theta_0)},
\end{equation}
where $\theta_0$ is the inclination angle of the observer with respect to $Z$-axis of the black hole. The contour of the shadow describes a circle
 $\alpha^2+\beta^2= \chi+\ell^2$. Note that the circular shape of the shadow boundary is true as long as the spacetime is spherically symmetric and static with any functional form of the metric functions $\mathscr{A}( \text{r} )$ and $\mathscr{B}( \text{r} )$, provided photon spheres and shadows do exist for such metric functions. The actual observational appearance of the shadow on observer's sky may differ from circularity due to the rotation of the black holes and the effect of the presence of the surrounding accretion disk. In  \cite{Tiede:2022bdd}, the authors analysed the appearance of the shadows of M87* as elliptical, based on the rotation of the black holes and the data sets produced from GRMHD simulations. However, we restrict our discussion here to non-rotating black holes only, for simplicity and also as the observed shadows do not seem to show a large deviation from circularity.  Therefore, the radius of the shadow is given by

\begin{equation}\label{sr}
    \begin{aligned}
    \text{r}_{\textbf{sh}} = \Bigg( \frac {1}{2}  \, \frac{ \mathscr{A}^{'}(\text{r}_{\textbf{ph}}) }{ \mathscr{A}^{2}( \text{r}_{\textbf{ph}} ) } \, \text{r}^{3}_{\textbf{ph}}  \Bigg)^{\frac{1}{2}} = \frac{ \text{r}_{\textbf{ph}} }{ \sqrt{ \mathscr{A}
    (\text{r}_{\textbf{ph}}}) }
    \end{aligned}
\end{equation}
The equation of photon sphere is
\begin{equation}\label{psg}
    \begin{aligned}
    2 \, \text{r}_{\textbf{ph}}^4 + 3 \,  \mathcal{P} \, \text{r}_{\textbf{ph}}^3 + 4 \, \mathcal{Q} \,  \text{r}_{\textbf{ph}}^2 + 5 \, \mathcal{S} \, \text{r}_{\textbf{ph}} + 6 \, \mathcal{T}  = 0
    \end{aligned}
\end{equation}

The condition for the existence of a horizon is given from the requirement $\text{g}_{\text{tt}}=0$ which leads to:
\begin{eqnarray}
r_\textbf{hr}^4 + {\cal P} r_\textbf{hr}^3 + {\cal Q} r_\textbf{hr}^2 + {\cal S} r_\textbf{hr} + {\cal T} =0
\end{eqnarray}

\noindent We will now rewrite the photon sphere equation and the horizon condition using dimensionless quantities mentioned before, i.e.
\begin{eqnarray}
{\cal P} = - 2\text{M} \hspace{0.2in};\hspace{0.2in} 
{\cal Q} = q \text{M}^2 \hspace{0.2in};\hspace{0.2in}
{\cal S} = s \text{M}^3 \hspace{0.2in};\hspace{0.2in}
{\cal T} =  t \text{M}^4
\end{eqnarray}
We further define
\begin{eqnarray}
x= \frac{\text{r}_\textbf{hr}}{\text{M}} \hspace{0.1in};\hspace{0.1in}
y= \frac{r_{\textbf{ph}}}{\text{M}}
\end{eqnarray}
With these redefinition the equations for $x$ and $y$ become:
\begin{eqnarray}
x^4- 2 x^3 + q x^2 + s x+ t=0,
\end{eqnarray}
and
\begin{eqnarray}
2 y^4 - 6 y^3 + 4 q y^2 + 5 s y + 6 t =0.
\end{eqnarray}
Both these equations now involve dimensionless quantities only.
From elementary algebra we know that the cubic terms in the quartic equations can be removed by simple transformations. If we write $x = z_1 +\frac{1}{2}$ and $y= z_2+ \frac{3}{4}$, the equations for $z_1$ and $z_2$ turn out to be
\begin{eqnarray}
z_1^4 + \left (q- \frac{3}{2}\right ) z_1^2 + \left (q+s-1\right )z_1 + 
\left (\frac{s}{2} +\frac{q}{4} + t -\frac{3}{16} \right ) = 0
\label{eq:z1}
\end{eqnarray}
and 
\begin{eqnarray}
z_2^4 + \left (2q-\frac{27}{8} \right )z_2^2 + \left (\frac{5s}{2} + 3 q -\frac{27}{8}\right ) z_2 +
\left (\frac{15s}{8} +\frac{9q}{8} + 3 t -\frac{243}{256}\right ) = 0
\label{eq:z2}
\end{eqnarray}
We will now explore some special case solutions of these equations
and obtain the expressions for the photon sphere, the horizon and the shadow radius as functions of one or more of the parameters $q$, $s$ and $t$. In general,
one may write
\begin{eqnarray}
r_{hr} = U(q,s,t) M \hspace{0.2in} ; \hspace{0.2in} r_{ph} = V(q,s,t) M
\hspace{0.2in} ; \hspace{0.2in} r_{sh} = W(q,s,t) M
\end{eqnarray}
where $U(0,0,0) = 2$, $V(0,0,0)=3$ and $W(0,0,0)=3\sqrt{3}$ and $M$ in dimensionful
form is $\frac{GM}{c^2}$. The general form of the three functions $U(q,s,t)$,
$V(q,s,t)$ and $W(q,s,t)$ can, in principle, be
obtained by solving for the real roots of the quartic equations quoted above.
We confine ourselves below to a study of special cases, largely
dictated by (a) the requirements of simplicity and (b) obvious mathematical
conditions which lead to simpler and tractable scenarios for which we
can find some viable answers.

\begin{figure}[!htbp]
\centering
\subfigure[$\frac{r_{hr}}{M}$(blue) and $\frac{r_{ph}}{M}$ (yellow) as a function of $t$ ($x$-axis).]{\includegraphics[width=3.2in,angle=360]{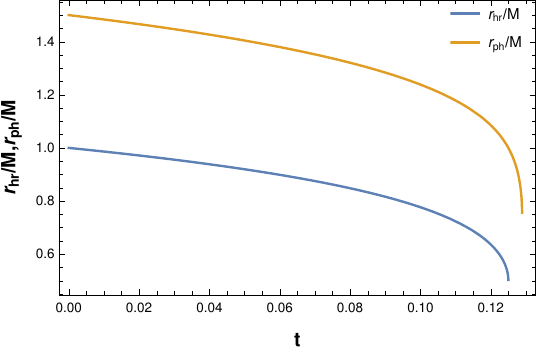}\label{subfig:fBa}}
\subfigure[Shadow radius $\frac{r_{sh}}{M}$ as a function of $t$.]{\includegraphics[width=3.2in,angle=360]{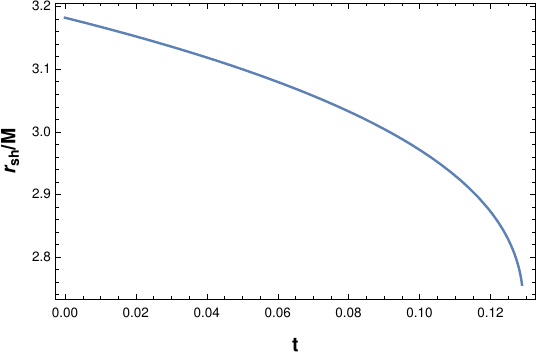}\label{subfig:fBb}}
\caption{\raggedright Horizon radius, photon sphere radius and shadow radius as a function of
$t$ for $q=7/4$, $s=-3/4$. Energy condition satisfying scenario for $t> \frac{25}{224}=0.111607$.
Note that the shadow radius is less than $3\sqrt{3}M$, i.e. lesser than Schwarzschild.}
\label{fig:B}
\end{figure}

\noindent {\bf Case 1:} This is the simplest case where we consider the
coefficients of $z_1$ and $z_2$ as zero. This results in algebraic
equations for $q$ and $s$ which have solutions
\begin{eqnarray}
q= \frac{7}{4} \hspace{0.2in} ; \hspace{0.2in} s = -\frac{3}{4}
\end{eqnarray}
We have seen that the energy conditions will hold provided 
${\cal S}^2 < \frac{72}{25} {\cal Q}{\cal T}$ which, with the
$q$, $s$, $t$ parametrisation becomes $s^2 < \frac{72}{25} q t$. With
$q=\frac{7}{4}$ and $s=-\frac{3}{4}$ this becomes $t> \frac{25}{224}$.
With this choice the equations for $z_1$, $z_2$ are
reduced to quadratic equations in $z_1^2$ and $z_2^2$. The real solutions
of these equations give
\begin{eqnarray}
z_1 = \sqrt{\frac{\sqrt{9-64 t}-1}{8}}\hspace{0.2in};\hspace{0.2in}
z_2 =\sqrt{\frac{2\sqrt{25-192 t}-1}{16}}
\end{eqnarray}
Recalling definitions, we finally obtain
\begin{eqnarray}
\text{r}_\textbf{hr} = \left (\frac{1}{2} + \sqrt{\frac{\sqrt{9-64 t}-1}{8}}\right ) \text{M}
\\
\text{r}_{\textbf{ph}} = \left ( \frac{3}{4} +  \sqrt{\frac{2\sqrt{25-192 t}-1}{16}}\right ) \text{M}
\end{eqnarray}
Notice that there is a restriction imposed on $t$ in order to have real values for $z_1$, $z_2$ and hence, $\text{r}_\textbf{hr}$ and $\text{r}_\textbf{ph}$. One can verify easily that within this domain $\text{r}_\textbf{hr} < \text{r}_\textbf{ph}$ (see Figure~\ref{subfig:fBa}). Note that for $t<0$ energy conditions are satisfied for $r>r_{+,max}$ as we noted earlier (see Table~\ref{table1}). One can verify that, for the specific choice of $q$ and $s$ here, the horizon radius $\text{r}_\textbf{hr}$ is always less than $r_{+,max}$, meaning there is always a violation of energy conditions outside the horizon up to the limit $r_{+,max}$.  Further, there is a range of $t>0$ (i.e. $0.061 <t< 0.1116$) over which one can still satisfy the energy conditions, at least outside the horizon of the spacetime. For $t\geq 25/224$ the energy conditions are satisfied for all $r$. However, $t> 1/8$ correspond to the naked singularities. 

\noindent Finally, we need to write down the shadow radius for any $t$.
This is a complicated expression, so we just plot it as a function of $t$ (see Figure~\ref{subfig:fBb}). Note that photon sphere and the shadow do not exist for the naked singularities for $t>99/768$.

%\begin{figure}[H]
%\begin{subfigure}{.5\textwidth}
%  \centering
%  \includegraphics[width=\linewidth]{case1-1.eps}  
%  \captionsetup{labelfont={bf,scriptsize},textfont={it,tiny}}
%  \caption{$\frac{r_{hr}}{M}$(blue) and $\frac{r_{ph}}{M}$ (yellow) as a function of $t$ ($x$-axis).}
%  \label{ec1}
%\end{subfigure}
%\begin{subfigure}{.5\textwidth}
%  \centering
%  \includegraphics[width=\linewidth]{case1-2.eps}  
%  \captionsetup{labelfont={bf,scriptsize},textfont={it,tiny}}
%  \caption{Shadow radius $\frac{r_{sh}}{M}$ as a function of $t$.}
%  \label{ec2}
%\end{subfigure}
%\caption{Horizon radius, photon sphere radius and shadow radius as a function of
%$t$. Energy condition satisfying scenario for $t> \frac{25}{224}=0.111607$.
%Note that the shadow radius is less than $3\sqrt{3}M$, i.e. lesser than Schwarzschild.}
%\end{figure}

\noindent The radius for a fixed $\text{M}$ is smaller than that for Schwarzschild, as is evident from the graph. We will need to choose other values of the 
free parameters in order to make it equal or larger than the \text{Schwarzschild}.

\

\noindent {\bf Case 2:} Another simplifying assumption
could be to assume the constant and the coefficients of the linear in the $z_1$, $z_2$ terms in the quartic equations to be zero.
Since we have three parameters $q$, $s$ and $t$ one needs three
equations. We can either assume both these coefficients in the
$z_1$ equation to be equal to zero and one of them in the $z_2$
equation to be zero. The opposite can also be considered. There will be four such choices given as:

\noindent (a) $q+s-1=0$, $8s+4q+16t=3$, $20s+24q-27=0$.

\noindent Here $q=\frac{7}{4}$, $s=-\frac{3}{4}$, $t=\frac{1}{8}$.
The energy condition requirement $s^2<\frac{72}{25} q t$ holds.
This corresponds to $t=\frac{1}{8}$ in Case 1 given above. 
Here $z_1=0$, hence $\text{r}_\textbf{hr} = 0.5 \text{M}$ and $\text{r}_\textbf{ph} = \text{M}$. The shadow radius can be read off from the figure.

\noindent (b) $q+s-1=0$, $8s+4q+16t=3$, $288q + 480s + 768 t -243=0$.

\noindent This set does not have a solution for $q$,$s$, $t$.

\noindent (c) $q+s-1=0$, $20s+24q-27=0$, $288q + 480s + 768 t -243=0$.

\noindent Here $q=\frac{7}{4}$, $s=-\frac{3}{4}$, $t=\frac{33}{256}$. The energy condition requirement holds.
The photon sphere here will be at $\text{r}_\textbf{ph} = 0.75 \text{M}$. There is
no horizon as the $z_1$ equation has no real root. Thus, we have
a naked singularity. The shadow radius is $\text{r}_\textbf{sh}=2.75568 M$. 
%\sout{One can find the shadow radius using the
%photon sphere radius.}

\noindent (d) $8s+4q+16t=3$, $20s+24q-27=0$, $288q + 480s + 768 t -243=0$.

\noindent Here $q=\frac{51}{32}$, $s=-\frac{9}{16}$, $t=\frac{9}{128}$. Here also the energy conditions hold 
good. 
%\sout{The only acceptable root of the $z_1$ quartic is
%$z_1=0$.} 
There are two real roots of the $z_1$ quartic equation, i.e. $z_1=0$ and $z_1=-0.0104$. Thus this describes a black hole with double horizon, where $z_1=0$ for the event horizon and $z_1=-0.0104$ for the cauchy horizon.
Thus, the event horizon is at $\text{r}_\textbf{hr} = 0.5 \text{M}$. The
photon sphere equation has the solutions $z_2=0$ and 
%\sout{$z_2= \frac{3}{16}$} 
$z_2=\pm \frac{\sqrt{3}}{4}$
%\sout{from which the photon sphere radii can be found}. 
One can verify that only $z_2=\frac{\sqrt{3}}{4}$ corresponds to the unstable photon orbit outside the event horizon. Thus the radius of the photon sphere is $\text{r}_\textbf{ph}=\frac{(3+\sqrt{3})M}{4}$ and the shadow radius $\text{r}_\textbf{sh}=3.11386 M$. 
%\sout{One can find the shadows using the photon sphere radius values.}

\

\noindent {\bf Case 3 (General):} We now go back to the quartic equations for $z_1$ and $z_2$, i.e. (\ref{eq:z1}) and (\ref{eq:z2}). Is it possible to arrive at some general analytical results on the roots? Following early work by Arnon
\cite{arnon} (see also another recent article \cite{prodanov}) we
are able to extract the following information which we state below.

\noindent Let us assume the quartic equation in the form
\begin{eqnarray}
u^4 + a u^2 + b u + c=0
\end{eqnarray}
We further define two quantities 
\begin{eqnarray}
\delta (a,b,c) =  256 c^3 - 128 a^2 c^2 + 144 a b^2 c + 16 a^4 c - 27 b^4 - 4 a^3b^2, \label{root cond1} \\
L(a,b,c) = 8 a c - 9 b^2 - 2 a^3.
\label{root cond2}
\end{eqnarray}
Real roots of the quartic will exist for specific conditions on
$\delta$ and $L$. In fact, the multiplicity of the real root can also
be obtained. The list appears in \cite{arnon} and \cite{prodanov}.
We mention the conditions of interest to us:

\noindent (a) The condition for no real roots is ($\delta >0$ and 
($L\leq 0$ or $a>0$)) OR ($\delta=0$ and $L=0$ and 
%\sout{$a=0$} 
$a>0$).

\noindent (b) The condition for two distinct real roots is just
$\delta<0$.

\noindent Six other conditions are stated in \cite{arnon} and \cite{prodanov}.

\noindent We will use (a) for $z_1$ to find examples of spacetimes 
without horizons, i.e. those with a naked singularity. We can use
(a) for $z_2$ to show there are no photon spheres and hence no shadow.
With (b) we can find horizons and photon spheres. In particular, if
we have one positive and one negative root we can use the 
obvious fact that $r>0$ in order to throw away the negative root.
That will leave us, from (b),  with one horizon and one photon sphere.

%\noindent Hence our strategy would be to make choices of $a$, $b$, $c$
%in accordance with our preference for a naked singularity or a black hole.
%Comparing the equation with that for $z_1$, we can extract $q$, $s$, $t$.
%Thereafter, we use the $q$, $s$, $t$ in the equation for $z_2$ and
%find the photon spheres and the corresponding shadow radius. Further,
%we need to check the energy conditions for the chosen $q$, $s$ and $t$.
%Let us now quickly look at some examples which are obtained following the
%above mentioned plan.

%The results appear in Table 1.
%\begin{table}
%\begin{center}
%\begin{tabular}{|c|c|c|c|c|c|c|c|}\hline \hline
%$(a,b,c)$ & $\delta$ & $(q,s,t)$ & $(z_1,z_2)$ & $r_H$ (in M)  & $r_{ph}$ (in M) & $r_{sh}$ (in M)
%& $s^2 < \frac{72}{25} q t ?$ \\ \hline \hline
%$(1,1,1)$ & $>0$ & ($\frac{3}{2},\frac{1}{2}, \frac{9}{16}$) & $(None, 0.2968)$ & $None $ & $1.047$ & $0.8968$ &  $Yes$ %\\ \hline 
%$(1,1,-1)$ & $<0$ & ($\frac{5}{2},-\frac{1}{2}, -\frac{19}{16}$) & $(0.5698, 1.6373)$ & $ 1.07 $ & $2.387$ & $3.458$ &  %$No$ \\ \hline

%\end{tabular}
%\end{center}
%\caption{Specific illustrative examples of the general case}
%\end{table}
%\vspace{0.2in}
%\centerline{Table I: Specific illustrative examples} 

\begin{figure}[!htbp]
\centering
\subfigure[]{\includegraphics[width=3in,angle=360]{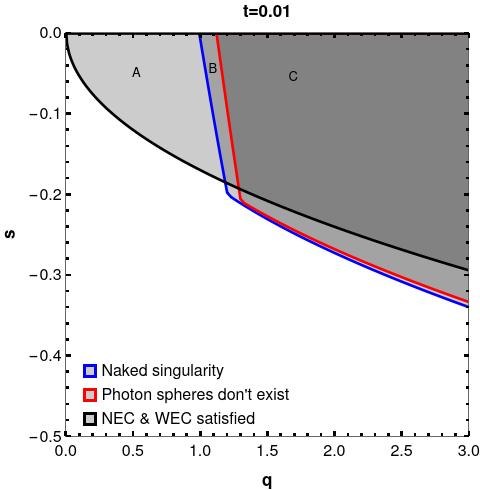}\label{subfig:f1a}}
\subfigure[]{\includegraphics[width=3in,angle=360]{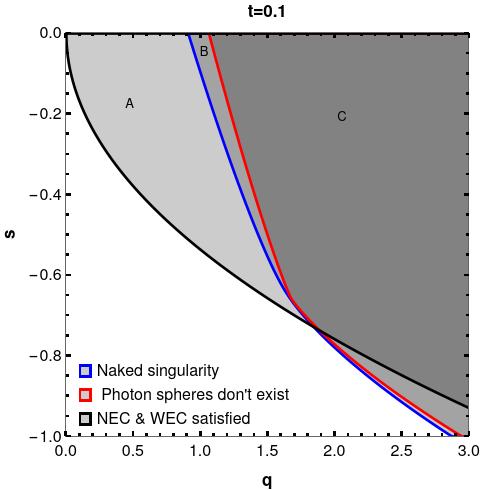}\label{subfig:f1b}}
\subfigure[]{\includegraphics[width=3in,angle=360]{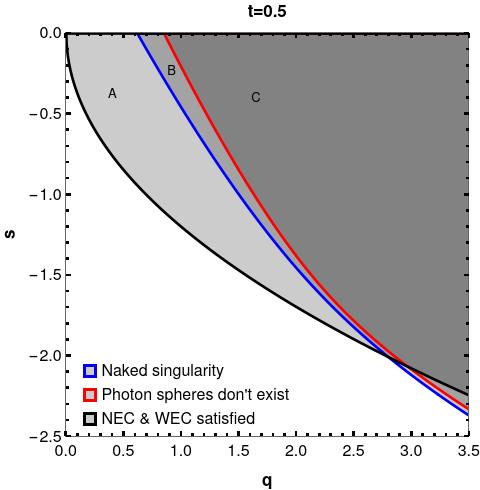}\label{subfig:f1c}}
\subfigure[]{\includegraphics[width=3in,angle=360]{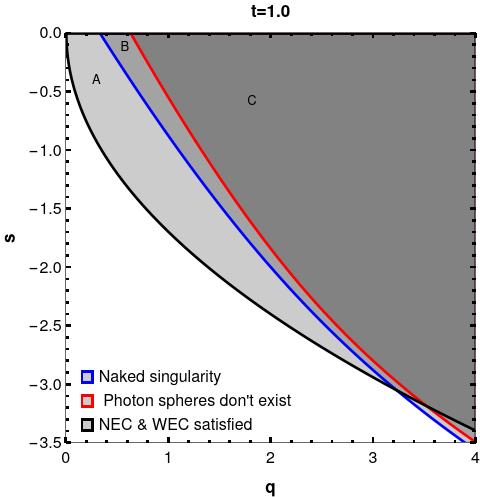}\label{subfig:f1d}}
\caption{\raggedright $q$ vs. $s$ region plot for fixed $t>0$. Note that we choose the domain $q>0$ and $s<0$. The shaded region with blue boundary correspond to naked singularities (the rest is for the black holes), the shaded region under the black boundary satisfies the energy conditions at all $r$, i.e. $s^2<\frac{72}{25}qt$, and for the shaded region with red boundary, photon spheres do not exist. The overlapping regions {\bf A} and {\bf B} correspond to allowed black holes and naked singularities respectively for which the energy conditions are satisfied and also the photon spheres exist. In the region {\bf C} the photon spheres do not exist though energy conditions are satisfied.}
\label{fig:C}
\end{figure}

\noindent In Fig.~\ref{fig:C}, we show $s$ vs. $q$ region plots for the chosen values of $t$. Each panel of the figure shows three shaded regions with uniform grey colour density but different boundary colours, i.e. black, blue, and red, signifying three different characteristics as mentioned in the figure. The region with black boundary satisfies the NEC and WEC, whereas the region outside do not satisfy the energy conditions. The shaded region under the blue boundary correspond to the naked singularities and the outside region is for the black holes. For the shaded region under the red boundary, photon spheres as well as the shadows do not exist. Note that the red boundary never crosses the blue boundary, meaning the non-existence of photon spheres only correspond to the naked singularities. However, there are overlaps of the parts of these three shaded regions--in the darker patches the overlap is more.
There are the overlapping regions {\bf A} for the black holes and {\bf B} for the naked singularities, where the photon spheres exist and also the energy conditions are satisfied. In {\bf C}, where parts of all the three shaded regions overlap, photon spheres and shadows do not exist although the energy conditions are satisfied.  As the $t$ values are increased as shown in the different panels of the figure, the region {\bf A} and {\bf B} become thinner in the direction of $q$ but lengthier along $s$, i.e. for a given choice of $q$ the allowed range of $s$ increases, but for a given choice of $s$ the allowed range of $q$ decreases. In Table~\ref{Table:II}, we demonstrate these facts with specific choices of $q$, $s$, and $t$. We take all possible combinations of $q=\lbrace \frac{1}{2}, \frac{4}{3} \rbrace$,   $s=\lbrace -\frac{1}{3}, -\frac{1}{10} \rbrace$, and  $t=\lbrace \frac{1}{100}, \frac{1}{10} \rbrace$ and compute the horizon radius, shadow radius, and check the energy conditions explicitly, thereby verifying their correspondence with the Fig.~\ref{fig:C}.

\begin{table}
\begin{center}
\begin{tabular}{|c|c|c|c|c|c|c|c|}\hline \hline
$(q,s,t)$ & $(\delta_1,L_1)$ & $(\delta_2,L_2)$ & $(z_1,z_2)$ & $r_H$ (in M)  & $r_{ph}$ (in M) & $r_{sh}$ (in M)
& $s^2 < \frac{72}{25} q t ?$ \\ \hline \hline
$(\frac{1}{2},-\frac{1}{10},\frac{1}{100})$ & $<0,\, <0$ & $<0,\, <0$ & $(1.24434, 1.90751)$ & $ 1.74434$ & $2.65751$ & $4.74945 $ &  $Yes$ \\ \hline 
$(\frac{1}{2},-\frac{1}{10},\frac{1}{10})$ & $<0,\, <0$ & $<0,\, <0$ & $(1.22413, 1.8909)$ & 1.72413 & 2.6409 & 4.73556 &  $Yes$ \\ \hline
$(\frac{1}{2},-\frac{1}{3},\frac{1}{100})$ & $<0,\, <0$ & $<0,\, <0$ & $(1.32444, 1.99484)$ & 1.82444 & 2.74484 & 4.83878 &  $No$ \\ \hline
$(\frac{1}{2},-\frac{1}{3},\frac{1}{10})$ & $<0,\, <0$ & $<0,\, <0$ & $(1.30854, 1.98082)$ & 1.80854 & 2.73082 & 4.82672 &  $Yes$ \\ \hline
$(\frac{4}{3},-\frac{1}{10},\frac{1}{100})$ & $>0,\, <0$ & $>0,\, <0$ & $(None, None)$ & $None$ & $None$ & $None$ &  $Yes$ \\ \hline
$(\frac{4}{3},-\frac{1}{10},\frac{1}{10})$ & $>0,\, <0$ & $>0,\, <0$ & $(None, None)$ & $None$ & $None$ & $None$ &  $Yes$ \\ \hline
$(\frac{4}{3},-\frac{1}{3},\frac{1}{100})$ & $<0,\, >0$ & $<0,\, >0$ & $(0.465475, 0.982887)$ & 0.965475 & 1.73289 & 3.63773 &  $No$ \\ \hline
$(\frac{4}{3},-\frac{1}{3},\frac{1}{10})$ & $>0,\, <0$ & $<0,\, <0$ & $(None, 0.775662)$ & $None$ & 1.52566 & 3.53269 &  $Yes$ \\ \hline
\end{tabular}
\end{center}
\caption{\label{Table:II} Specific illustrative examples of the general case. Each entry can be easily correlated with the Fig.~\ref{subfig:f1a} and Fig.~\ref{subfig:f1b} and the conditions on the Eqs.~(\ref{root cond1}) and (\ref{root cond2}); $(\delta_1, L_1)$ and $(\delta_2, L_2)$ correspond to the quartic equations of $z_1$ and $z_2$ respectively. }
\end{table}

%\newpage
\section{\bf Link with observational results from M87* and SgrA*} \label{section5}

\noindent It is known from the EHT observations that the shadow radius for the supermassive compact object (black hole) in the M87 galaxy \cite{event2019first} and Sgr A* (in the center of our galaxy) \cite{EHT2022_1}, correspond to angular diameters $42\pm 3$ $\mu$as and $48.7\pm 7$ $\mu$as, respectively. Assuming the Schwarzschild black hole model, we can obtain the shadow radius using the formula $r_{sh} = \frac{3\sqrt{3} GM}{c^2}$ for the given masses and distances of these objects. The angular diameter value for the black hole in M87 turns out to be $39.6$ $\mu$as (considering the mass $M=6.5\times 10^9\, M_{\odot}$ and the distance $D=16.8\, \text{Mpc}$ ). For Sgr A*, the Schwarzschild value is around $51.2$ $\mu$as (considering the mass $M=4\times 10^6\, M_{\odot}$ and the distance $D=8\, \text{Kpc}$).
We will now see if the $r_{sh}$ for the line element proposed here can
match the observations within error bars. In other words, we ask-- what values of
$q,s,t$ can yield a shadow radius close or equal to the value found in observations.
%\begin{figure}[H]
%\centering
% \includegraphics[width=0.7\linewidth]{m87.eps}  
  %\captionsetup{labelfont={bf,scriptsize},textfont={it,tiny}}
 % \caption{$q=t=-s$, $q$ negative. $q$ is on the x-axis. The y-axis represents
 % the shadow radius (blue), horizon radius (red) divided by $M$. 
 % The green and yellow
 % horizontal lines represents the range allowed for M87*, i.e. for angular diameter range $42\pm 3$ $\mu$as.}
 % \label{m87-1a}
%\end{figure}
%\begin{figure}[H]
%\centering
% \includegraphics[width=0.7\linewidth]{sgr.eps}  
  %\captionsetup{labelfont={bf,scriptsize},textfont={it,tiny}}
%  \caption{$q=t=-s$, $q$ negative. $q$ is on the x-axis. The y-axis represents
 % the shadow radius (blue), horizon radius (red) divided by $M$. 
 % The green and yellow
 % horizontal lines represents the range allowed for SgrA*, i.e. for angular diameter range $48.7\pm 7$ $\mu$as.}
 % \label{m87-1b}
%\end{figure}

\noindent For M87*, the angular diameter can be close to the observed value
provided the $3\sqrt{3}$ factor is replaced by numerical values ranging from
$4.94$ (lower bound $39$ $\mu$) as to $5.71$ (upper bound $45$ $\mu$as).
In the case of SgrA* the corresponding values are: 4.1 (for 
$41.7$ $\mu$as) and $5.46$ (for $55.7$ $\mu$as).
We show below that there does exist ($q, s, t)$ values for
which one can safely say that observations match with theory. 

\noindent We first consider the case with
%\sout{$q$ negative and $s=t=-q$} 
$s$ negative and $q=t=-s$. 
The energy conditions obviously hold good without any further restrictions.
The roots $z_1$, $z_2$ are found as functions of $q$ and thereafter
$r_{hr}$, $r_{ph}$ and $r_{sh}$ are also obtained as functions of $q$.
The expressions are long and complicated and are therefore not quoted
here. However, one can check that $q\leq 1.1169$ for black holes, otherwise they are naked singularities. 

\begin{figure}[!htbp]
\centering
\subfigure[]{\includegraphics[width=3.1in,angle=360]{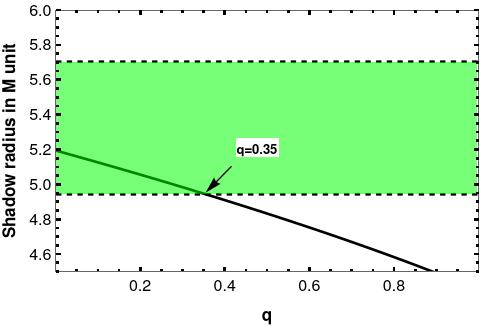}\label{subfig:fDa}}
\subfigure[]{\includegraphics[width=3.1in,angle=360]{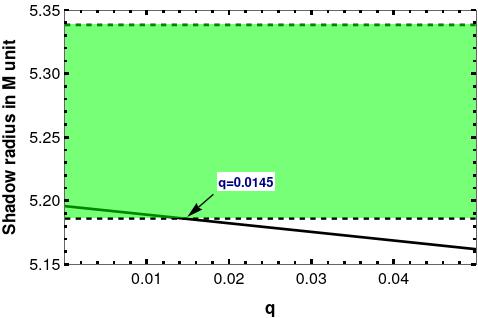}\label{subfig:fDb}}
\caption{\raggedright Shadow radius ($r_{sh}$) in $M$ units vs. $q$ plot (black solid curve) for the case $q=-s=t$. The green band is the observational range for the black hole M87$^*$. In the plot $(a)$, we take earlier version of the angular shadow diameter range $42\pm 3$ $\mu \text{as}$. In the plot $(b)$, we take the new updated (PRIMO) version of the range  $41.5\pm 0.6$ $\mu \text{as}$. The dashed lines are the upper and lower limits of the observational range.   }
\label{fig:D}
\end{figure}

\begin{figure}[!htbp]
\centering
\subfigure[]{\includegraphics[width=3.1in,angle=360]{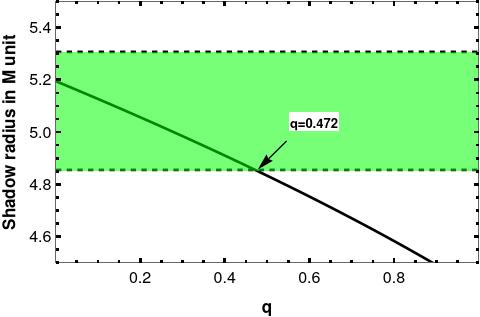}\label{subfig:fEa}}
\subfigure[]{\includegraphics[width=3.1in,angle=360]{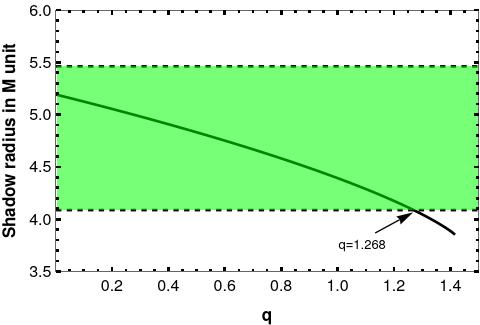}\label{subfig:fEb}}
\caption{\raggedright Shadow radius ($r_{sh}$) in $M$ units vs. $q$ plot (black solid curve) for the case $q=-s=t$. The green band is the observational range for the black hole SgrA$^*$. In the plot $(a)$, we take the angular (emission) ring diameter range $51.8\pm 2.3$ $\mu \text{as}$. In the plot $(b)$, we take the  angular shadow diameter range  $48.7\pm 7$ $\mu \text{as}$. The dashed lines are the upper and lower limits of the observational range. }
\label{fig:E}
\end{figure}

\noindent The figures below (Fig.~\ref{fig:D} and Fig.~\ref{fig:E}) show the possibilities and the $q$ values for which
a match with M87* and SgrA* observations is feasible. We note, 
%\sout{from the graphs} 
that the horizon radius is always smaller than the photon sphere radius.
%\sout{Also, there is only one photon sphere. The corresponding geometry is a black hole with a
%horizon.} 
The metrics in general are:

%\begin{eqnarray}
%\textcolor{red}{g^{rr}=-g_{tt}= \left (1-\frac{2M}{r} - \frac{\vert q\vert  M^2}{r^2} +
%\frac{\vert q \vert M^3}{r^3} - \frac{\vert q \vert M^4}{r^4}\right )}
%\nonumber
%\end{eqnarray}

\begin{eqnarray}
g^{rr}=-g_{tt}= \left (1-\frac{2M}{r} + \frac{ q  M^2}{r^2} -
\frac{q M^3}{r^3} + \frac{q M^4}{r^4}\right )
\label{metric q=-s=t}
\end{eqnarray}
One can identify the above line element with a Kiselev black hole with appropriately chosen $K_i$.
The range of $q$ in the graphs is chosen to avoid any jumps in the curves which may occur due to
pathologies in the roots for certain $q$ values. 
%\sout{The allowed $\vert q\vert$ values are
%the ones for which the shadow radius lies within the green and yellow horizontal lines.} 
The allowed $q$ values are the ones for which the shadow radius lies in the observational green band in the figures.
%\sout{For example $q=-1$ is a valid choice for which the shadow radius will match with M87 observations. 
%It is important to note that the same
%$q$ value (say $q=-0.4$) may describe both observations .}

\noindent For the black hole M87$^*$, observational limit is $q\leq 0.35$ (see Fig.~\ref{fig:D}(a)) obtained from the earlier version of the observational range of the angular shadow diameter. However, the new improved (PRIMO) version \cite{Medeiros_2023} of the range of angular diameter puts a more stringent limit $q\leq 0.0145$.

\noindent For the black hole SgrA$^*$, the EHT papers \cite{EHT2022_1,EHT2022_6} provide two observational data for the angular diameter. One is mentioned as emission ring diameter $51.8\pm 2.3$ $\mu \text{as}$ and the other one is mentioned as the shadow diameter $48.7\pm 7$ $\mu \text{as}$. We consider both as the observed shadow diameter and obtain the constraints on the $q$ parameter. From Fig.~\ref{fig:E}.(a) we find  that $q\leq 0.472$ 
whereas $q\leq 1.268$ from  Fig.~\ref{fig:E}.(b). The latter bound on $q$ implies the possibility of SgrA$^*$ being a naked singularity as well. We scrutinize this possibility further using the constraint on fractional deviation parameter ($\delta$) which signals a deviation from the Schwarzschild 
black hole. 

\begin{figure}[!htbp]
\centering
 \includegraphics[width=0.5\linewidth]{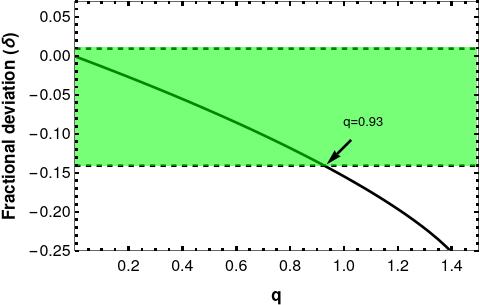}  
  %\captionsetup{labelfont={bf,scriptsize},textfont={it,tiny}}
  \caption{$q=-s=t$, $q>0$. $q$ is on the x-axis. The y-axis represents
  the fractional deviation parameter $\delta$. The black solid line is the theoretical. The green band bounded by lower and upper dashed lines is observational range for the black hole SgrA$^*$ from VLTI and Keck bounds.}
  \label{fig.F}
\end{figure}

\noindent The recent EHT papers on SgrA$^*$ observations have used the fractional deviation parameter $\delta$ to constrain the spacetime geometries different from the Schwarzschild or Kerr black holes. The definition of $\delta$ is as follows,
\begin{equation}
    \delta = \frac{d_{sh}}{d_{sh,Sch}}-1 =\frac{R_{avg}}{3\sqrt{3}M} -1 ,
\end{equation}
 where the average diameter of the shadow, $d_{sh}=2R_{avg}$. Using the observations of the shadow of SgrA$^*$ and two separate sets of prior values of mass and distance of SgrA$^*$ from the VLTI and Keck observations, the EHT  collaboration provided the bound on $\delta$ \cite{EHT2022_1,EHT2022_6} as
 \begin{equation}
 \delta= \begin{cases} 
      -0.08^{+0.09}_{-0.09}& \text{(VLTI)} \\
      -0.04^{+0.09}_{-0.10} & \text{(Keck)}  \\
   \end{cases}
\end{equation}
Therefore, combining both these bounds, we have $-0.14 <\delta < 0.01$. Using this constraint we obtain observational limit $q\leq 0.93$ (see Fig.~\ref{fig.F}). Considering all the obtained bounds on $q$ together, we get the best possible one, $q\leq 0.472$, for the black hole SgrA$^{*}$, with
the present accuracy of available data. However, it is expected 
that improved values will appear when the PRIMO technique is applied to the data for SgrA$^*$.  

%\noindent In order to facilitate a comparison with the metric
%that arises in Rastall gravity we must try and see what happens if
%either the term in $\frac{1}{r^3}$  or the term in $\frac{1}{r^4}$ is nonzero. For the term $\frac{1}{r^3}$  we look at the case $t=0$ but $s \neq 0$, and, for the term $\frac{1}{r^4}$,  we look at  $s=0$ but $t \neq 0$. We scan the parameter space $q$ vs $s$ and $q$ vs $t$ respectively for the case $\frac{1}{r^3}$ and $\frac{1}{r^4}$.   In Figs.~\ref{fig:Ga}, \ref{fig:Gb},~\ref{fig:Ha},~\ref{fig:Hb}, we show the contour plots in the parameter space. The allowed regions are specified by the contour lines marked with the observational values. One can see from the metric in Eq.~(\ref{Rastall}) that for $\kappa \lambda=\frac{2}{3}$ one retains the term
%in $\frac{1}{r^3}$ and for $\kappa \lambda=\frac{1}{2}$ one retains the term
%in $\frac{1}{r^4}$ . Thus the shadow corresponding to the metric in Rastall gravity
%can also be modeled with a suitable choice of parameters in
%a special case of our general line element. 
\begin{figure}[!htbp]
\centering
\subfigure[ M87$^*$ (old angular diameter)]{\includegraphics[width=3in,angle=360]{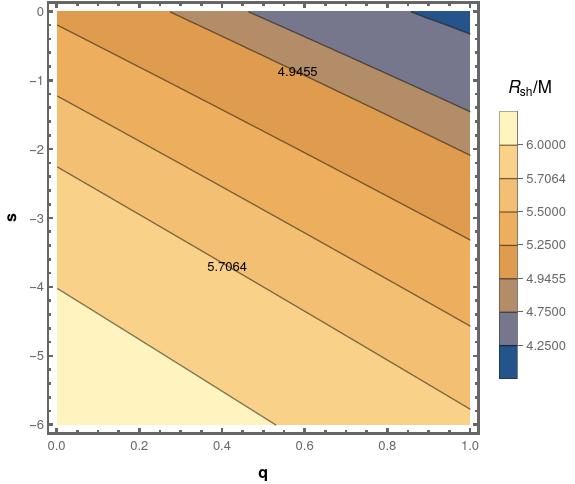}\label{subfig:fGa}}
\subfigure[M87$^*$ (new angular diameter)]{\includegraphics[width=3in,angle=360]{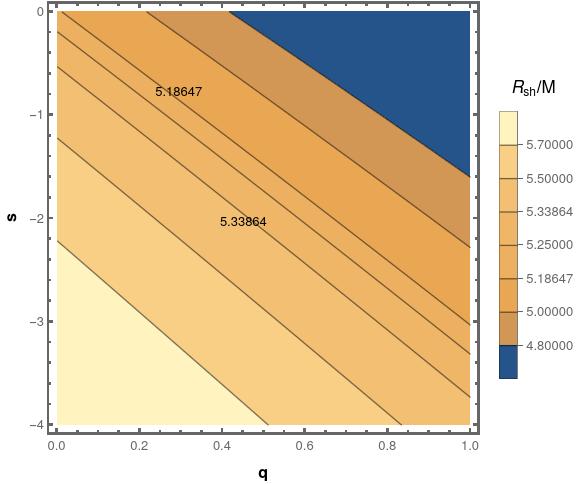}\label{subfig:fGb}}
\caption{\raggedright \footnotesize $q$ vs. $s$ contour plots for the case $t=0$ (for M87*). The area bounded by the marked contour lines are only allowed in each plot. In the plot $(a)$, we take earlier version of the angular shadow diameter range $42\pm 3$ $\mu \text{as}$ for the M87*. Contours represent the fixed values of the shadow radius in the mass unit. The observational limits on the shadow radius are marked on the specific contours in the plot.  In the plot $(b)$, we take the new updated (PRIMO) version of the range  $41.5\pm 0.6$ $\mu \text{as}$ for the M87*.
}
\label{fig:Ga}
\end{figure}

\begin{figure}[!htbp]
\centering
\subfigure[ SgrA* (ring diameter)]{\includegraphics[width=3in,angle=360]{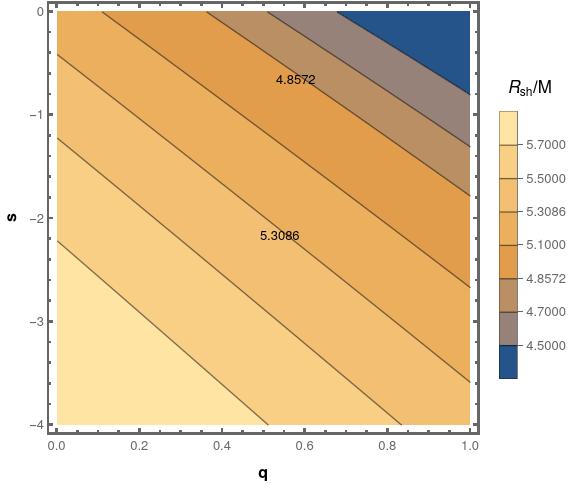}\label{subfig:fGc}}
\subfigure[SgrA$^*$ (fractional deviation parameter)]{\includegraphics[width=3in,angle=360]{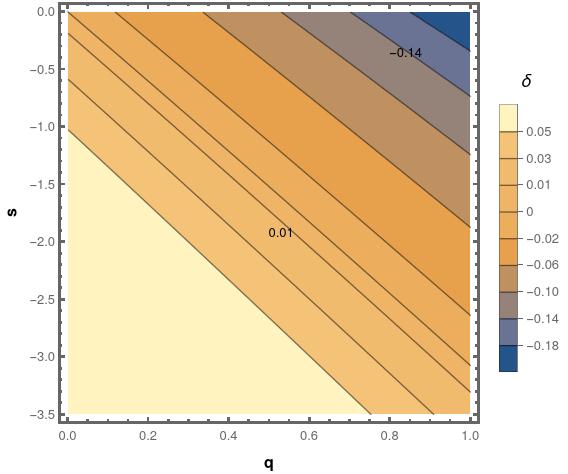}\label{subfig:fGd}}
\caption{\raggedright \footnotesize  $q$ vs. $s$ contour plots for the case $t=0$ (for SgrA*). 
In the plot $(a)$, the angular (emission) ring diameter range $51.8\pm 2.3$ $\mu \text{as}$ is considered for the SgrA* . In the plot $(b)$, we use the observational range for the fractional deviation parameter ($\delta$) for the SgrA*. Here, the contours represent the fixed values of $\delta$ with the observational ones are marked on the plot.}
\label{fig:Gb}
\end{figure}

\begin{figure}[!htbp]
\centering
\subfigure[ M87$^*$ (old angular diameter)]{\includegraphics[width=3.2in,angle=360]{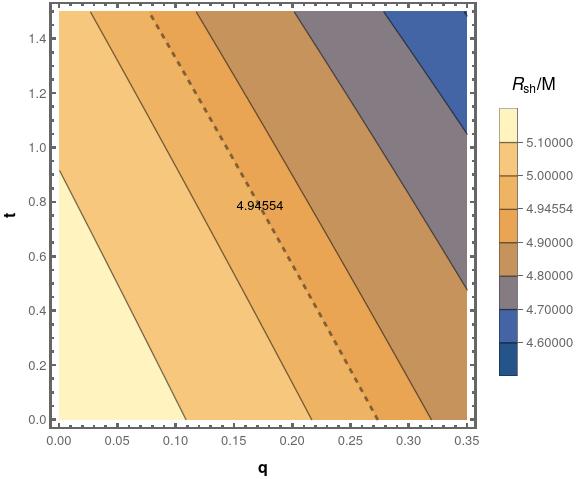}\label{subfig:fHa}}
\subfigure[M87$^*$ (new angular diameter)]{\includegraphics[width=3.2in,angle=360]{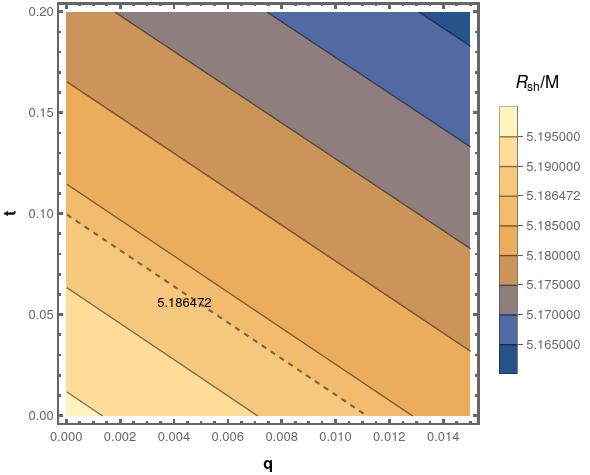}\label{subfig:fHb}}
\caption{\raggedright  $q$ vs. $t$ contour plots for the case $s=0$ (for M87*). 
}
\label{fig:Ha}
\end{figure}

\begin{figure}[!htbp]
\centering
\subfigure[ SgrA* (ring diameter)]{\includegraphics[width=3.2in,angle=360]{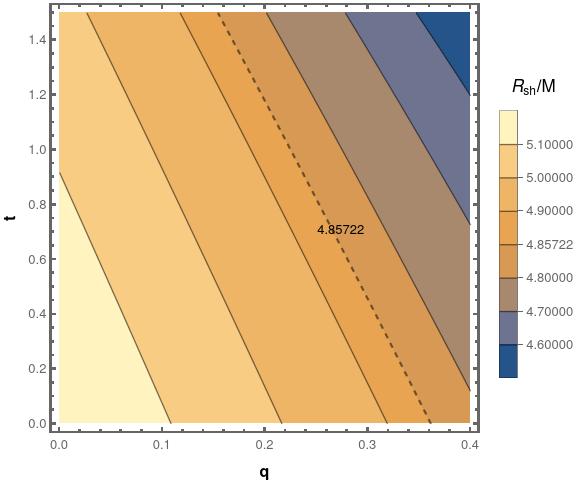}\label{subfig:fHc}}
\subfigure[SgrA$^*$ (fractional deviation parameter)]{\includegraphics[width=3.2in,angle=360]{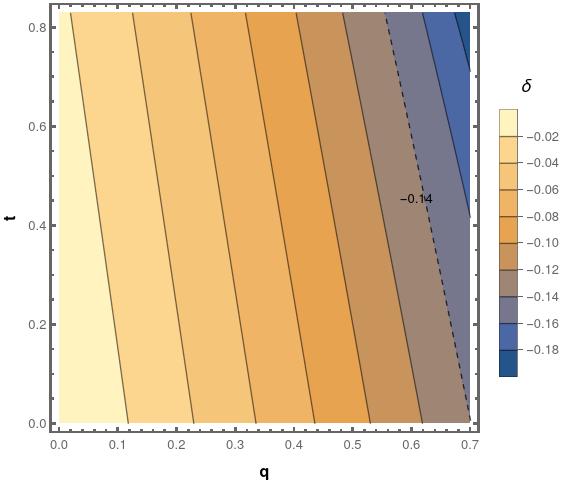}\label{subfig:fHd}}
\caption{\raggedright  $q$ vs. $t$ contour plots for the case $s=0$ (for SgrA*). }
\label{fig:Hb}
\end{figure}
%\newpage

 In order to facilitate a comparison with the metric
that arises in Rastall gravity we must try and see what happens if
either the term in $\frac{1}{r^3}$  or the term in $\frac{1}{r^4}$ is nonzero. For the term $\frac{1}{r^3}$  we look at the case $t=0$ but $s \neq 0$, and, for the term $\frac{1}{r^4}$,  we look at  $s=0$ but $t \neq 0$. We scan the parameter space $q$ vs $s$ and $q$ vs $t$ respectively for the cases with  $\frac{1}{r^3}$ and $\frac{1}{r^4}$.  In Figs.~\ref{fig:Ga}, \ref{fig:Gb},~\ref{fig:Ha},~\ref{fig:Hb}, we show the contour plots in the parameter space. The allowed regions are specified by the contour lines marked with the observational values. One can see from the metric in Eq.~(\ref{Rastall}) that for $\kappa \lambda=\frac{2}{3}$ one retains the term
in $\frac{1}{r^3}$ and for $\kappa \lambda=\frac{1}{2}$ one retains the term
in $\frac{1}{r^4}$ . Thus the shadow corresponding to the metric in Rastall gravity
can also be modeled with a suitable choice of parameters in
a special case of our general line element.

 It is also possible to consider more general cases where all of $q$, $s$, and $t$ take non-zero values. For example, one can choose a specific value of $t$ and then analyze the parameter space $q$ vs. $s$ for the allowed region.

\section{A more general metric and consequences} \label{section6}

\noindent Let us now consider a more general metric (arbitrary but finite $N$) for which the $\text{g}_\text{tt}$
and $\text{g}_\text{rr}$ are given as:
\begin{eqnarray}
\text{g}_\text{tt} = -\left (1 +\sum_{n=1}^{N} \frac{\text{a}_\text{n}}{\text{r}^\text{n}}\right ) = -\frac{1}{\text{g}_\text{rr}}
\end{eqnarray}
Here $\text{a}_\text{n}$ are the generalisations of the ${\cal P}, {\cal Q}, {\cal R}, {\cal T}$
discussed just above. 

\noindent We begin with the photon sphere equation in this metric which turns out to be
\begin{eqnarray}
2 \, \text{r}_\textbf{ph}^\text{N} + \sum_{\text{n}=1}^\text{N} \text{a}_\text{n} (\text{n}+2) \, \text{r}_\textbf{ph}^\text{N-n} = 0
\label{ph sph eqn gen}
\end{eqnarray}
On the other hand the horizon equation for the general case is 
\begin{equation}
    \text{r}_\textbf{hz}^\text{N} + \sum_{n=1}^N \text{a}_\text{n} \text{r}_\textbf{hz}^{\text{N-n}} =0
    \label{hz eqn gen}
\end{equation}
Below we discuss two special solutions of the general case.\\
$(a)$ First we assume a special choice wherein the general photon sphere equation~(\ref{ph sph eqn gen}) 
reduces to the form 
\begin{eqnarray}
\left (\text{r}_\textbf{ph} - \text{p} \right )^\text{N} = \text{r}_\textbf{ph}^\text{N} - \text{N} \, \text{r}_\textbf{ph}^{\text{N}-1} \, \text{p} + .....+ (-1)^\text{N} \,  \text{p}^\text{N} = 0.
\end{eqnarray}
This means that the photon sphere is located at $\text{r}_\textbf{ph} = \text{p}$. Of course this will happen only for a specific set of $\text{a}_\text{n}$ in the original metric.
Comparing the two photon sphere equations (i.e. 6.2 and 6.4) one finds that
\begin{eqnarray}
\text{a}_\text{n} = (-1)^\text{n} \, \frac{2\text{N}!}{\text{n}! \, (\text{N-n})! \, (\text{n}+2)} \, \text{p}^\text{n}
\end{eqnarray}
The general metric now takes on a special form with
\begin{eqnarray}
\text{g}_\text{tt} = - \left (1 + \sum_{\text{n}=1}^\text{N} \frac{2 \, (-1)^\text{n} \,  \text{N}!}{\text{n}! \, (\text{N-n})! \, (\text{n}+2)} \left (\frac{\text{p}}{\text{r}}\right )^ \text{n}\right ) = -\frac{1}{\text{g}_\text{rr}}
\end{eqnarray}
It is now easy to write down the shadow radius $\text{r}_\textbf{sh}$, which, after some
straightforward algebra becomes
\begin{eqnarray}
\text{r}_\textbf{sh} = \sqrt{\frac{(\text{N}+1)(\text{N}+2)}{2}} \, \text{p}
\end{eqnarray}

\noindent   One can 
check that the two inequalities $\rho\geq 0$ and  $\rho + \tau_{\text{i}}\geq 0$ for satisfying the WEC and NEC, reduce to
the following:
\begin{eqnarray}
\sum_{\text{n}=1}^\text{N} \frac{2 \, (-1)^{\text{n}+1} \, (1-\text{n}) \, \text{N}!}{\text{n}! \, (\text{N-n})! \, (\text{n}+2)} \frac{\text{p}^\text{n}}{\text{r}^{\text{n}+2}} \geq 0
\\
\sum_{\text{n}=1}^\text{N} \frac{(-1)^{\text{n}+2} \, (\text{n}-1) \, \text{N}!}{\text{n}! \, (\text{N-n)}!} \frac{\text{p}^\text{n}}{\text{r}^{\text{n}+2}} \geq 0 
\end{eqnarray}

\noindent Thus, one will need to check the energy conditions for different N (or, if possible for a general \text{N} directly) and also find out if the spacetime has horizons and where the horizon is located. Let us quickly check what happens for
$\text{N}=4$. For this case, we find
\begin{eqnarray}
\text{g}_\text{tt} = - \left ( 1-\frac{8 \text{p}}{3\text{r}} + \frac{3 \text{p}^2}{\text{r}^2} -\frac{8 \text{p}^3}{5 \text{r}^3} +\frac{\text{p}^4}{3 \text{r}^4} \right )
\end{eqnarray}
A straightforward calculations shows that there are no horizons and the geometry represents a naked singularity. The shadow radius is given as:
\begin{eqnarray}
\text{r}_\textbf{sh} = \sqrt{15} \, \text{p} 
\end{eqnarray}
The energy condition inequalities lead to the relations
\begin{eqnarray}
\frac{15\text{r}^{2} \text{p}^{2} -16 \text{p}^{3} \text{r} + 5 \text{p}^{4}}{\text{r}^{6}} \geq 0 \\
\frac{6 \text{p}^{2} \text{r}^{2} - 8 \text{p}^{3} \text{r}+ 3 \text{p}^{4}}{\text{r}^{6}} \geq 0
\end{eqnarray}
One may verify (use $\frac{\text{r}}{\text{p}}$ as a variable) that both these inequalities hold good
and the energy conditions are satisfied for all $\text{r}$. 

\noindent One can define $\text{p} = \nu \frac{\text{GM}}{\text{c}^{2}}$ in order to bring in the right dimensions and the mass dependence. 
It is also instructive to look at what happens for $\text{N}=1$--this is the simple case of the vacuum \text{Schwarzschild} spacetime for which
\begin{eqnarray}
\text{g}_\text{tt} = - \left ( 1-\frac{2\text{p}}{3 \text{r}}\right ) = -\frac{1}{\text{g}_\text{rr}}
\end{eqnarray}
Choosing $\text{p}=\frac{3 \text{GM}}{\text{c}^2}$ one ends up with the standard form of the metric
and the shadow radius
also reduces to the well-known $3 \sqrt{3} \frac{\text{GM}}{\text{c}^2}$.

\noindent $(b)$ We obtain another special solution by making a choice for the form of the general horizon equation~(\ref{hz eqn gen})
\begin{equation}
    (\text{r}_\textbf{hz}- \text{e})^\text{N}=0,
    \label{extremal gen}
\end{equation}
where the horizon equation has the only root `$\text{e}$' (here e is not the Euler number, but any arbitrary positive quantity) with multiplicity $\text{N}$. Then the coefficients of the general metric take the form
\begin{equation}
    \text{a}_\textbf{n}= (-1)^\text{n}\frac{\text{N!}}{\text{n!}\text{(N-n)}!}\text{e}^\text{n},
\end{equation}
and the resulting metric components are
\begin{equation}
    \text{g}_{\text{tt}}=-\left(1-\frac{\text{e}}{\text{r}}\right)^\text{N}=-\text{g}_{\text{rr}}^{-1}.
\end{equation}

\noindent The spacetime represents a generic version of the extremal Reissner-Nordstr\"om (RN) black holes. For $\text{N}=2$ we recover a RN black hole with mass $\text{M}=\text{e}$ and the charge $Q=\pm \text{e}$ (note that $\text{e}$  must be positive for a black hole). For $\text{N}=4$, the metric component takes the form
\begin{equation}
    \text{g}_{\text{tt}}=-\left(1-\frac{4\text{e}}{\rr}+\frac{6\text{e}^2}{\rr^2}-\frac{4\text{e}^3}{\rr^3}+\frac{\text{e}^4}{\rr^4}\right),
\end{equation}
where the mass of the black hole is then $\text{M}=2\text{e}$ and the effective charge $Q=\pm \sqrt{6}\text{e}$.
The shadow radius for such a class of extremal black holes is given by
\begin{equation}
    {\text{r}}_{\textbf{sh}}=\text{e} \left(1+\frac{\text{N}}{2}\right) \left(1+\frac{2}{\text{N}}\right)^{\frac{\text{N}}{2}}.
\end{equation}
The validity of NEC and WEC leads to the inequality
\begin{equation}
\left[1+\frac{\text{e}}{\rr-\text{e}}\right]^{\text{N}}-\left[1+\frac{\text{e}\text{N}}{\rr -\text{e}}\right]\geq 0,
\end{equation}
which is satisfied for all $\text{r}\geq \text{e}$, i.e. outside the horizon.

\section{Conclusions} \label{section7}
\noindent Schwarzschild and Reissner-Nordstr\"om spacetimes are the simplest and most widely used representatives of spherically symmetric, static black holes. In this paper, we have considered a more general static, spherisymmetric spacetime in the Schwarzschild gauge,
by considering a general series involving $\frac{1}{r^k}$ ($k\geq 0$) terms, with arbitrary constant coefficients, in the metric functions. Such a metric may be a solution in GR coupled with matter fields or may be a solution in a modified/ alternative theory of gravity.  
It could also represent black holes or other real or hypothetical
objects in astrophysics. We have first analyzed the energy conditions (NEC and WEC) 
for the matter required (assuming the Einstein field equations of GR), if such a general spacetime has to exist. Thereafter, we restrict the parameter space 
of our line element by comparing the size of the circular shadow profile with observational data for M87$^*$ and SgrA$^*$.

    \noindent Our generic metric  (\ref{metg}) is parametrized by four (assuming terms upto $\frac{1}{r^4}$) constants ${\cal P},\, \cal{Q},\, \cal{S},\, \cal{T}$, where the effective mass and charge of the black hole/ naked singularity are related as ${\cal P}=2\text{M}_{\textbf{eff}}$ and $\text{Q}_{\textbf{eff}}=\pm\sqrt{{\cal Q}}$. Analysis of the energy conditions reveal that NEC and WEC are satisfied for all $r$ for any of the following two conditions only: $(i)$ ${\cal Q}, {\cal S}, {\cal T}>0$, and $(ii)$ ${\cal Q}>0$, ${\cal S}<0$, ${\cal T}>0$, and ${\cal S}^2\leq \frac{72}{25}{\cal Q}{\cal T}$. 
 The first condition implies a naked singularity only and, for the second condition, the spacetime may be a black hole or a naked singularity.   For any other choice of the parameters the energy conditions are partially or completely violated. The details are provided in Table~\ref{table1} and Fig.~\ref{fig:A0}. 

 \noindent Next we study the critical null geodesic (i.e. photon sphere) and the shadow. We obtain a quartic horizon equation~(\ref{eq:z1}) and a quartic photon sphere equation~(\ref{eq:z2}) for the general case. By a root analysis, we scan the parameter space to identify viable regions for black holes/naked singularities and the existence of photon spheres, along with the energy conditions (see Fig.~\ref{fig:C} and the Table~\ref{Table:II} for illustration). We also demonstrate some special solutions of the horizon and photon sphere equations. 

 \noindent Thereafter, we compare the observed angular diameters of the M87* and SgrA* with the theoretical prediction for different parameter values satisfying the energy conditions. In particular, we choose $q=-s=t$ (for which the metric takes the form ~(\ref{metric q=-s=t}) ). For this metric,  comparing with all observations, we get a bound on $q\leq 0.0145$ for M87* (with updated PRIMO data) and $q\leq 0.472$ for SgrA*. We also scan $q$ vs. $s$ parameter space ($t=0$ for $1/r^3$ term) in Figs.~\ref{fig:Ga},~\ref{fig:Gb} and $q$ vs. $t$ parameter space ($s=0$ for $1/r^4$ term) in Figs.~\ref{fig:Ha},~\ref{fig:Hb} for observationally allowed regions. Similar plots may also be obtained for more general cases where $q,\,s,\,t\neq 0$.

\noindent Finally, we try to further generalize the metric functions by including higher order terms $\frac{1}{r^N}$ where $N$ is arbitrary but 
integral, positive, 
and finite. However, for arbitrary $N$ 
and the corresponding constant coefficients, 
any known simple solution with viable matter fields is not really known. We obtain two special solutions for which either the photon sphere equation or the horizon equation has a single root with multiplicity $N$. In the first case, 
for $N=4$, the solution turns out to be naked singularity whereas in the second case, the solution (for $N=4$) seems to bear a resemblance (via $M=2\text{e}$) 
to an extremal Reissner-Nordstr\"om black hole but with higher powers
of $\frac{1}{r}$ included.  

 \noindent In the section~II, we provided some specific examples which 
 are indeed special cases of the general metric quoted in Eqn. (\ref{metg}). In particular, we noted how such special cases of the general metric may represent black holes or naked singularities surrounded by some fluids, namely the Kiselev type solutions both in GR and modified gravity or other solutions
 as known in diverse contexts. As mentioned before, there do exist
 many  more line elements which fall within the broad class assumed in
 Eqn. (\ref{metg}). Thus, our generic results on energy conditions and shadows may
 help in understanding features associated with specific line elements 
 which have arisen in 
 different theories of gravity, over the years.  It may however be necessary to extend our analysis here, if further higher order terms in $\frac{1}{r}$ 
 have to be considered in the metric functions. Inclusion of such terms will
 not change the circular shape of the shadow -- it will only make the
 equations more complicated (higher degree polynomials).

 \noindent  Finally, one possible and natural extension of our work is to include rotation, i.e. construct
 a line element with rotation. One way to do this is to employ the Newman-Janis 
 method \cite{Newman:1965tw,Newmn:1965b} of generating rotating spacetimes from static ones. 
 Once we have the rotating spacetimes we can proceed in the same way, as done in this article and construct the shadow profiles. In such a scenario though,
 the shadow profiles will be much more complicated (no longer simple circles) and it is likely
 that one will have to use numerical methods to arrive at any meaningful 
 conclusions. We hope to return to this aspect as well as other, similar questions in our future endeavours.  

\section*{Acknowledgements}
\noindent Research of SJ is partially supported by the SERB, DST, Govt. of India, through a TARE fellowship
grant no. TAR/2021/000354, hosted by the Department of Physics, Indian Institute of Technology Kharagpur.

\section*{Data Availability Statement}
\noindent The manuscript has no data associated with it.

\bibliography{shadowref}

\end{document}